\documentclass[a4paper,12pt]{article}
\usepackage{amsfonts}
\usepackage{amssymb}
\usepackage{amsmath}
\usepackage{epsfig}
\def\uqa{U_q(\widehat{sl(2)})}

\def\beq{\begin{equation}}
\def\eeq{\end{equation}}
\def\bea{\begin{eqnarray}}
\def\eea{\end{eqnarray}}

\def\u1{\widehat{U(1)}}
\def\su2{\widehat{SU(2)}_1}

\def\a{\alpha}
\def\b{\beta}

\def\l{\lambda}

\def\de{\partial}

\def\nn{\nonumber} 

\def\flecha {\rightarrow}
\def\Uqa{\widehat{U_q(sl(2))}}
\def\epsi{\epsilon}

\catcode`\@=11
\def\numberbysection{\@addtoreset{equation}{section}
        \def\theequation{\thesection.\arabic{equation}}}

\numberbysection

\begin{document}

\begin{titlepage}

\begin{center}
\hfill  \quad  \\
\vskip 1 cm
{\LARGE \bf Effective Field Theory and Integrability in }
\vspace{0.5 cm}
{\LARGE \bf  Two-Dimensional Mott Transition}

\vspace{1cm}

\vspace{.4cm}
Federico~L.~ BOTTESI ,\ \ Guillermo~R.~ZEMBA\footnote{
Fellow of CONICET, Argentina.}\\

\bigskip
{\em Facultad de Ingenier\'ia  Pontificia Universidad Cat\'olica Argentina,}\\
{\em  Av Alicia Moreau de Justo 1500, 1428, Buenos Aires, Argentina\\}
\medskip
{\em Physics Department,}\\
{\em  Comisi\'on Nacional de Energ\'{\i}a At\'omica,} \\
{\em Av.Libertador 8250, (1429) Buenos Aires, Argentina}

\end{center}
\vspace{.5cm}
\begin{abstract}
\noindent
We study the Mott transition in a two-dimensional lattice spinless fermion model with 
nearest neighbors density-density interactions. By means of a two-dimensional Jordan-Wigner 
transformation, the model
is mapped onto the lattice $XXZ$ spin model, which is shown to possess a Quantum Group symmetry
as a consequence of a recently found solution of the Zamolodchikov Tetrahedron Equation. 
A projection (from three to two space-time dimensions) property of the solution is 
used to identify the symmetry of the model at the Mott critical 
point as $\uqa \otimes \uqa$, with deformation parameter $q=-1$. 
Based on this result, the low-energy Effective Field theory for the model
is obtained  and shown to be a lattice double Chern-Simons theory with coupling constant 
$k=1$ (with the standard normalization). 
By further employing the Effective Filed Theory methods, we show that the Mott transition that arises 
is of topological nature, with vortices in an 
antiferromagnetic array and matter currents characterized by a $d$-density wave order parameter. 
We also analyze the behavior of the system upon weak coupling, and conclude that
it undergoes a quantum gas-liquid transition which belongs to 
the Ising universality class.
\end{abstract}

\vskip 1.cm


\vfill
\hfill October 2010
\end{titlepage}
\pagenumbering{arabic}

\section{Introduction}\label{section-1}

In spite of many substantial advances in recent years, the study of some problems in the physics of strongly correlated electrons continues to provide stimulating challenges. Among the several reasons for this, we could mention the still incomplete understanding of all of the properties of strongly correlated electron systems and the lack of many reliable techniques to study them. 
One of these paradigmatic problems is the Mott transition, loosely defined as a metal-insulator transition driven by correlations. As early as in 1939, Mott argued that if the electronic density in metallic systems was lowered enough, the Coulomb repulsion would dominate over the kinetic energy and the system could undergo a transition to an insulating state \cite{Mott}.

On the one hand, many widely employed theoretical descriptions of the Mott transition are based on the study of the microscopic dynamics of the electronic system: one starts by writing down a model Hamiltonian for the electrons, sometimes coupled to external fields o to some other degrees of freedom,  and then tries to solve it within some approximation or through the use of numerical methods.  However, numerical methods like exact diagonalization are restricted to small clusters, and if the interaction among electrons is strong enough, the typical approximation schemes based on the resummation of some class of Feynman diagrams are not completely reliable. For these reasons, it is a difficult task to find solutions displaying the Mott transition, even in the simplest models, like the Hubbard one.  As of today,  there exist two successful approaches for study this transition based on microscopic dynamics: one is the dynamical mean field theory (DMFT) method \cite{Kotliar-DMFT}, valid in the limit of infinite spatial dimensions, which neglects  the  spatial correlations; the second approach consists in finding analytic expressions for physical observables in integrable models (mainly in one-dimensional systems) by means of the Bethe Ansatz or bosonization methods.              
  
On the other hand, systems like conventional superconductors or quantum Hall systems have universal properties that are well described by field theories which do not deal with the microscopic degrees of freedom, but rather with fields representing {\it effective} degrees of freedom. These two cases are classic examples of the more general framework of the Effective Field Theories (EFT) approach, which has its roots in Landau's ideas for condensed matter systems and which is widely and
successfully used in high energy physics. This approach can be considered as 'way of thinking' which emphasizes the symmetries of the systems and that naturally incorporates Wilson's renormalization group ideas. In this approach, 
the study of a system starts by wisely choosing the {\it effective degree of freedom}, which are the relevant ones at a given energy scale, and then one proceeds to write down the most general second quantized action compatible with the characteristic symmetries of these degrees of freedom, retaining only the marginal and relevant terms, {\it i.e.}, terms 
that are non-decreasing in the low energy (long-distance) limit \cite{Polchinski}.            
During the last years, several exotic states that contain droplets of approximately constant density have been found experimentally in electronic systems considered to be close to the Mott transition. These states that may seem to be surprising and difficult to explain from the point of view of the free electrons, are good candidates to be understood 
following the EFT approach. For example, the appearance of effective gauge forces arising from the dynamics, and
which have not been included in the microscopic electron Hamiltonian can be properly taken into account within the EFT framework\cite{Laughlin}. 

The goal of this article is to construct and consider an EFT for a two-dimensional square lattice system
which displays the Mott transition, which implies that we shall focus our attention on the symmetry aspects of this
transition.
Specifically, we will consider a simple model of electrons with nearest neighbors density-density interaction 
which has also been previously studied, with the goal of identifying its effective degrees of freedom and 
their characteristic symmetries. Since the scope of this article is to make it readable to both 
Condensed Matter and Field Theory physicists, we shall also review (without pretending to be exhaustive) 
some basic aspects of bosonization, Conformal Field Theory, and Integrable Models.   

The paper is organized as follows: 
in Section 2, we present the model of strongly coupled fermions on a (two-dimensional) square lattice
that we shall consider in the paper. We review some known properties of the corresponding one-dimensional 
version of this model and we also discuss the relationship among different approaches for treating the one-dimensional Mott transition.  We also apply a two-dimensional bosonization prescription as considered in \cite{Fradkin} for the two-dimensional fermion model. 
After reviewing some basic properties of integrability in statistical mechanics models, we discuss in the Section 3 the integrability of the specific two-dimensional fermion model considered in this paper. 
We show that in the strong coupling regime, the system defined by the ground and low-lying states of the model 
satisfies the Zamolodchikov Tetrahedron Equation, and is characterized by a novel family of solution to the 
Tetrahedron Equation recently found by Bazhanov et. al. \cite{Bazhanov-1} \cite{Bazhanov-2}. We review
these solutions for the sake of completeness and discuss the three-dimensional structure of an
underlying Quantum Group algebraic structure. This analysis allows us to identify the symmetry of the model at the 
Mott transition point as given by the Quantum Group  $\Uqa \otimes \Uqa$. The identification of the symmetry and the corresponding effective degrees of freedom allows us to write down the EFT for the model, which is done in the Section 4. With the EFT at hand, we then analyze the order parameter and the universality class of the transition. We find that it 
is given by a Kosterlitz- Thouless type transition, with vortices in an anti-ferromagnetic array. We also discuss how this order is modified by doping, and that this procedure induces an Ising-like phase transition. Finally, we present our Conclusions.     
\pagebreak 

\section{The two-dimensional fermion lattice model}\label{section-2}

In this Section we introduce the model that we shall be considering throughout the paper.
Let us consider a spinless fermions system with nearest neighbors interaction on a square lattice, with Hamiltonian
\beq
H_{2d}\ =\ -\frac{t}{2}\ \sum_{x,\mu} [\ \psi^\dagger(x+ a e_\mu) e^{i A_\mu} \psi(x)\ +\ {\rm h.c.}\ ]\  +\ U\ \sum_{x,\mu} \rho(x)\rho(x+a e_\mu)\ , \label{Model-Ferm-2d}
\eeq
where $\psi(x)$ is the fermionic field, $x$ labels the lattice sites and $e_\mu$ are the unit lattice vectors pointing to the nearest neighbors of a given site, $a$ is  the lattice spacing , $t$ is the hopping parameter, $U$ is the (constant) Coulomb potential, $\rho(x)$ is the  charge density (normal-ordered with respect to the half-filling ground state),  $\rho(x)= [:\psi^{\dagger}(x)\psi(x): -1/2]$ and $A_\mu $ is an Abelian statistical gauge field defined on the links of the lattice.


\subsection{Review of the one-dimensional model} 

In order to proceed in our study of this model,
we first would like to review the physics of the one-dimensional model analog of (\ref{Model-Ferm-2d}),
given by the Hamiltonian
\beq
H_{1d}\ =\ -\frac{t}{2}\ \sum_{x} [\ \psi^\dagger(x+a)  \psi(x)\ +\ {\rm h.c.}\ ]\  +\ U\ \sum_{x} \rho(x)\rho(x+a)\ , 
\label{Model-Ferm-1d}
\eeq 
were the sums are taken over the lattice sites. 
Note that the gauge field is unimportant in this case, as it should, given  
that there are no statistical Gauge fields in one spatial dimension (see,{\it e.g.}, \cite{Fradkin-book}). 
This model has an interesting history that begins with the work of Luther and Peschel \cite{Luther-Peschel} and which has later on been studied in detail by 
several authors, including Shankar\cite{Shankar}. 
The Mott transition in one-dimensional systems has been 
discussed not only in the context of Hamiltonian models like (\ref{Model-Ferm-1d}), 
but also within the scope of Luttinger liquids. 
In the following, we review and relate both of these approaches 
with the scope of setting up a framework suitable for
further generalizations and for finding 
the effective degrees of freedom for the simplest one-dimensional case.   

As it is well-known, the model (\ref{Model-Ferm-1d}) 
can be mapped onto the (one-dimensional) $XXZ$ model
through the Jordan-Wigner transformation:
\begin{eqnarray}
&& S^+(i)=\psi^\dagger _i \exp{( i\pi\sum_{i<j}\ \psi^\dagger_j\psi_j)}\ , \\
&& S^-(i)= \exp{( -i\pi\sum_{i<j} \psi^\dagger_j\psi_j)}\  \psi_i\ , \\
&& S_z(i)=\psi^\dagger(i) \psi(i)-1/2\ ,
\end{eqnarray}
where $i$ labels the lattice sites. The Hamiltonian goes onto
\beq
H^{1d}_{xxz}\ =\ \sum_i\ [\ -(S^x_iS^x_{i+1}+S^y_iS_{i+1})\ +\ \Delta S^z(i) S^z(i+1)\ ]\label{hxxz}\ .
\eeq
In the thermodynamical limit, 
when the total number of sites is even, this Hamiltonian corresponds to a the
system at half-filling, which has the property: 
\beq
H_{xxz}(\Delta)\ =\ -H_{xxz}(-\Delta) \label{delta-delta}\ .
\eeq
Moreover, the $XXZ$ model is integrable and its spectrum and
other analytic properties have been found in
\cite{Wu-Lieb} using the Bethe Ansatz. Its ground state energy is:

\begin{eqnarray}
E_0=\begin{cases}
\frac{1}{4}\cosh\lambda -\ \frac{1}{4} \sinh \l\ [\l+2\l\sum_n(1+e^{2n\l })^{-1}\ ] & \text{if $\Delta=\cosh \l > 1$}\\
1/4-\ln 2 &\text{if $\Delta=1 $} \\
\frac{1}{4} \cos \mu-\sin^2 \mu \int_{- \infty} ^\infty dx/[2 \cosh \pi x(\cosh 2 \mu x-\cos\mu  )]& 
\text{ if $\Delta=\cos\mu <1$}
\end{cases}
\end{eqnarray}
Note that $E_0$ is an analytic function of $\Delta$  in the range 
$1<\Delta <\infty$, so that the singularity at $\Delta=1$ signals a phase 
transition. This fact has been used in \cite{Shankar} to show that
the transition at at $\Delta=1$ is identified as a Mott one.
The argument relies on the duality between two
opposite regimes for the system, ranging from   
the insulator behavior for $\Delta \to \infty$ to 
the metallic one for $\Delta=0$. 
Moreover, it is known that for $\Delta=1+\epsilon$ the spins are in a 
Neel state, and therefore the system must be in a charge density wave (CDW)
state. 

An alternative description of the system (\ref{Model-Ferm-1d}) is given 
by bosonization of its fermionic degrees of freedom. 
For $\Delta < 1$, the action of the system is given by
\bea
S=\frac{g}{4\pi} \int dz  d\bar{z}\ \de_z  \phi(z)\  \de_{\bar{z}}\bar{\phi}(\bar{z})\ ,
\label{Free-Boson-action} 
\eea
where $g$ is a self-coupling parameter and we have defined complex space-time coordinates $z=x+it$ and $\bar{z}=x-it$, 
and the normal ordered charge density is given by $\rho(z)=i\de_z \phi(z)$ . 
The model has effective degrees of freedom  which are bosonic fields representing charge density waves.
Eq. (\ref{Free-Boson-action}) defines a conformal field theory (CFT) 
whose energy-momentum tensor has holomorphic and anti-holomorphic components given by:
\bea
T(z) &=&-g\de_z \phi(z) \\  
\bar{T}({\bar z}) &=&-g\de_{\bar{z}} \bar{\phi}(\bar{z})\ .\eea
Moreover, the Fourier modes of the fields, defined by 
\bea
&& T(z)=\sum_n L_n z^{-n-2}\ , \\
&& \rho(z)=\sum_n \rho_n z^{-n-1}\ , \eea
satisfy the following chiral algebra:
\bea
&&[\ L_n,L_m\ ] = (n-m)\ L_{n+m }\ +\ \frac{c}{12}\ \delta_{n+m,0}\ (n^3 - n) \\
&&[\ \rho_n , \rho_m\ ] = n\ \delta_{n+m,o} \\
&&[\ L_m,\rho_n\ ] = -m\ \rho_{n+m}\ . 
\eea
These three lines define a current algebra:
the first line is the Virasoro algebra for 
the generators $L_n$ with central charge $c=1$. 
The second one is the $\widehat{u(1)}$ current (or Kac-Moody) algebra
for the charge modes. The third is required for
consistency among the other two. The $\widehat{u(1)}$ current algebra  
is used to define the Luttinger model, 
which in Hamiltonian form is usually written as: 
\beq
H\ =\ \frac{1}{2\pi }\int dx \left[u K(\pi \Pi(x)
)^{2}+\frac{u}{K}\left(\partial _{x}\phi \right)^{2}\right]\ ,
\label{1d-bosonized-2} 
\eeq 
where $u$ , $K$  are called the Luttinger parameters. 
It is straightforward to regain the action (\ref{Free-Boson-action})
with coupling constant $g=K$
starting from the Hamiltonian (\ref{1d-bosonized-2}), 
by transforming the arguments of the fields into imaginary time. The parameters of the model in the different representations are related by (for a  detailed discussion see \cite{Schultz}):
\beq
K=\frac{\pi}{2[\pi-\arccos(\Delta)]}
\eeq  
Hence, the Mott transition in the one-dimensional lattice fermion system 
(\ref{Model-Ferm-1d}) is characterized by $\Delta=1$ or $K=g=1/2$. The 
Luther-Emery transformation \cite{Luther-Emery} allows us to rewrite the 
Hamiltonian (\ref{1d-bosonized-2}) as $H = H_0+H_1$, where:
\bea
&& H_0=v \int   p\ [\psi^\dagger_+ (p)\psi_+(p)- \psi^\dagger_-(p) \psi_-(p)]\ dp  \\
&& H_1=\frac{\pi  u }{L}\sinh(2\theta)[\int( 2\rho_+ \rho_-  +f_1\sum_{\a =\pm} :\rho_\a(p)\rho_\a(-p):) dp\ ]\ ,
\label{Lutter-Emery-ponit}
\eea
where $ v=u(\cosh 2\theta + f_1 \sinh 2\theta ) $, $\exp(2\theta)=1/(2K)$ and $f_1$ is an arbitrary constant. Note that for the Luther-Emery line, which coincides with Mott transition, {\it i.e.}, $K =1/2$, one has that $H_1$ vanishes so that the theory consists of two free decoupled chiral fermions. 
Therefore, we identify the effective degrees of freedom of the theory at the Mott transition as two free fermionic currents of opposite chirality.


Next we would like to discuss the characteristic symmetry of the 
degrees of freedom and, therefore, of the system at Mott transition point. 
In order to do this, it would be more convenient to switch to the spin representation ({\it i.e.}, the $XXZ$ model). 
Note that the relation (\ref{delta-delta}) allows us to write
\beq
H^{1d}_{xxz}\ =\ \sum_i \left[ (S^x_iS^x_{i+1}+S^y_iS_{i+1})\ -\ \frac{(q+q^{-1})}{2} S^z_i S^z_{i+1}\label{Spin-Chain} \right] \ ,
\eeq
where $ q=\exp{i\gamma}$, $\gamma=\cos \Delta$ 
{\it i.e.}, $\Delta= -(q+q^{-1})/2$. For $q=-1$, we have 
the isotropic anti-ferromagnetic spin chain, which is a critical system 
with an explicit $SU(2)$ symmetry.  

It is well-known that the corresponding low energy effective field theory 
is the Wess-Zumino-Witten (WZW) model with coupling constant 
$k=1$  \cite{Afleck} and action:
\beq
S_{WZW}\ =\ \frac{k}{4\pi} \int d^2x\ {\rm Tr}[\de_\mu g \de ^\mu g^{-1}]\ +\ \frac{k}{12\pi} \int_{B(D)} d^3x Tr[\epsilon ^{\mu \nu \lambda } g^{-1} \de_\mu g g^{-1} \de_\nu g g^{-1} \de_\lambda g] \ , 
\label{wzwa}
\eeq   
where $g(z)$ is a field in the group manifold of $SU(2)$ ({\it i.e.}, $g$ is a $SU(2)$-valued
matrix). The first integral in (\ref{wzwa}) is defined over a compactified two-dimensional domain $D$ and the second is done over a three-dimensional ball with boundary $D$. For the sake of completeness, we will sketch here the procedure leading to this EFT. Following \cite {Afleck}, one makes a 
transformation from the spin variables $S_i$ to a fermionic system $\psi_i^\a $ that preserves the 
$SU(2)$ symmetry defined by 
${\bf S}_i=1/2\sum_{\a,\b}\psi_i^\a {\bf \sigma} \psi_i^\b $ ,where ${\bf \sigma}=(\sigma_x,\sigma_y,\sigma_z)$ and $\sigma_i$ are the Pauli matrices. The standard commutation relations reproduce the correct spin commutators, however the Hilbert space of the fermion system is too large and one must restrict it by projecting out the states with one particle by site 
\beq 
(\psi_n^\dagger)^\a (\psi_n)_\a=1\ . \label{vinculo}\eeq
In the low energy regime, the only excitations that should be taken into account are localized around the two Fermi points of the one-dimensional
Fermi surface. One transforms to a set of new fermionic degrees of freedom  
(we will label the lattice sites by the integer $n$ to avoid confusion):
\beq
\psi_\a(n) =\ \sqrt{a}\left[ i^n \psi_{\a L}(n\pm 1/2)+ (-i)^n \psi_{\a R}(n\pm 1/2) \right ]\ .
\eeq
A final redefinition of variables takes us to current operators: 
\bea
&& J=i :(\psi_L^\dagger)^\a (\psi_L)_\a: \qquad  J^i=i :(\psi^\dagger_L )^\a \sigma ^i (\psi_L)_\a : \\  
&& G=\psi^\dagger_L \psi_R\ +\ \psi^\dagger_R \psi_L \qquad  G^i=i :\psi^\dagger_L\sigma ^i \psi_R:\  
+\ :\psi^\dagger_R \sigma ^i \psi_L:\ ,
\eea
where $R$ and $L$ denote the left and right chiral components. The spin operators and the 
constraint (\ref{vinculo}) become:
\bea
&& J+\bar{J}=G=0\label{vinculo-2}\\
&& S^i/a=J^i+\bar{J}^i+(-1)^n G^i \ .
\eea
In the continuum limit, the Hamiltonian (\ref{Spin-Chain}) of the spin chain becomes:
\bea
H=\frac{a}{2}\ \int d^2x\ \left[\ J^i(x)J^i(x)\ +\ \bar{J}^i(x)\bar{J}^i(x)\ +\ 
2\ J^i\bar{J}^i(x)\ \right]\ , 
\label{hcont}
\eea
where the $x$ variable is the continuum limit of the 
lattice position. 
The last term in (\ref{hcont}) is irrelevant in the renormalization group sense
and the effective Hamiltonian becomes the Hamiltonian of the WZW model.


\subsection{Bosonization of the two-dimensional fermion model} \label{section-3}

We now turn our attention back to the two-dimensional fermionic model on the square lattice
with Hamiltonian (\ref{Model-Ferm-2d}). It can be bosonized as discussed, {\it e.g.}
in \cite{Fradkin-2DJW } and \cite{Tsvelik}, by applying a two-dimensional Jordan-Wigner transformation. 
Let us first consider the case $U=0$. As it is known from the one-dimensional 
case, the Jordan-Wigner transformation owns its existence to a 
natural ordering of the particles along the line. 
This ordering is lost in on two-dimensional lattice, but the mapping could
still be defined by adding extra degrees of freedom, in the form of attached statistical fluxes 
to the particles, 
{\it i.e.}, by introducing branch-cuts on the otherwise analytic fermionic field operators. 
Equivalently, one considers the Hamiltonian (\ref{Model-Ferm-2d}) with the additional Gauss law constrain:
\beq
\rho(x)-\theta B(r)=0 \ ,
\label{constrain} 
\eeq
where $\rho(x)$ is the charge density and $B(r)$ is the magnetic field 
defined on sites of the dual lattice, {\it i.e.}, a lattice obtained from the original (direct) one by translating 
its set of vertices to the centers of each plaquette of the direct lattice:
\bea
&& B=\epsilon_{ij}\Delta_i A_j \\
&& \Delta_i A_j=A_j(x+e_i)-A(x) 
\eea
The Gauss law constraint (\ref{constrain}) implies that for each fermion on the site 
$x$ of the direct lattice, there is also a quantum flux (or vortex) in the 
corresponding site $r$ of the dual lattice. It can 
be implemented at the field theory level by coupling the 
fermions to an Abelian statistical Chern-Simons (CS) Gauge field. 
In order to show how to do it, let us consider the Lagrangian: 
\bea
L_{2d}  &=& \sum_{x}\ \psi^\dagger(x)iD_0 \psi(x)\ - t\sum_{x,j=1,2}\ [\psi^\dagger(x) e^{iA_j}\psi(x+e_j)\ + 
{\rm hc}\ ]\nn \\ 
&& +\ \frac{\theta}{4}\ \sum_x \epsilon_{\mu,\nu,\l}\  A^{\mu}(x)F^{\nu \l}(x)\ , \eea
where 
\bea
&& D_0=\de_0 -iA_0\\
&& F_{ij}=\Delta_iA_j-\Delta_j A_i\\
&& F_{0i}=\de_0 A_i- \Delta_i A_0 \ .\eea
The canonical quantization of the above Lagrangian
in the Gauge $A_0=0$ imposes the constraint (\ref{constrain}) at the level of the 
Hilbert space \cite{Dunne-Jackiw-Trugenberger}.
The classical solutions of the constraint can be written 
as follows:
\beq
A_j\ =\Delta_j\phi(x)=  \frac{1}{\theta}\sum_{x'}[\Theta(x+e_j,r')-\Theta(x,r')]\rho(x')\ , \eeq
where $\Theta(x,r)$ is the lattice-angle function \cite{Luscher} which contains a branch cut 
from $r$ to $\infty$ satisfying:
$\oint_\Gamma \Delta \theta =+1 $
for any closed curve $\Gamma$ that encloses the point $r$ of the dual lattice. 
Moreover, the operators:
\bea
&& a(x)=e^{i\phi(x)} \psi(x)\label{bosop-1} \\
&& a^\dagger(x)=\psi^\dagger(x) e^{-i\phi(x)} \label{bosop-2}
\eea
satisfy bosonic commutations relations for $1/2\theta=m\pi$ and due to the Pauli principle, the operator (\ref{bosop-1}) are hard-core bosons representing charge density waves
(CDW) states of the underlying electrons.So that the mapping defined by the equations (\ref{bosop-1}) (\ref{bosop-2}), relate the fermionic degree of freedom to bosonic ones, and therefore the spinless fermion system can be mapped onto
the following Hamiltonian:
\beq
H^{2d'}_{Bos}\ =\ -t\sum_{x,\mu=1,2} \left[ a^\dagger(x)a(x+e_\mu)\ +\ {\rm h.c.} \right]
\ +\ U\sum_x \rho(x)\rho(x+a)\ .
\eeq
The identifications:
\bea
&& S^+(x)=a^\dagger(x) \\
&&  S^-(x)=a(x) \\
&& S^z(x)=\rho(x)-1/2 \label{bosop-3}\eea
where the operators $S^+(x)$ and $S^-(x)$ are the raising and lowering spin operators for a $s=1/2$ spin particle, allows us to map the original fermion system onto the two-dimensional $XXZ$-spin model:
\beq
H^{2d}_{XXZ}\ =\ \sum_{\langle i j\rangle}\left [ -(\ S^x_i S^x_j\ +\ S^y_i S^y_j\ ) +\Delta S^z_i S^z_j \right]
\eeq
where $i,j$ denote lattice sites and the sum is taken over nearest neighbors , and  we  have defined $\Delta=t/U$

Summarizing, fermionic  two-dimensional systems can be bosonized  by attaching fluxes to particles, which is achieved at quantum level by constraining the Hilbert space states after imposing the Gauss Law. The dynamics of these
systems may be described at the Hamiltonian level in terms of these bosonic degrees of freedom,  which
physically represent charge density waves. Alternatively, 
the charge density waves may be replaced by another set of degrees of freedom, such as  
the spin waves in the so-called $XXZ$ spin model.  

\section{Integrability of the two-dimensional model}

\begin{figure}[h!]  
\centering
\resizebox{15.cm}{!}{ 
\includegraphics[clip]{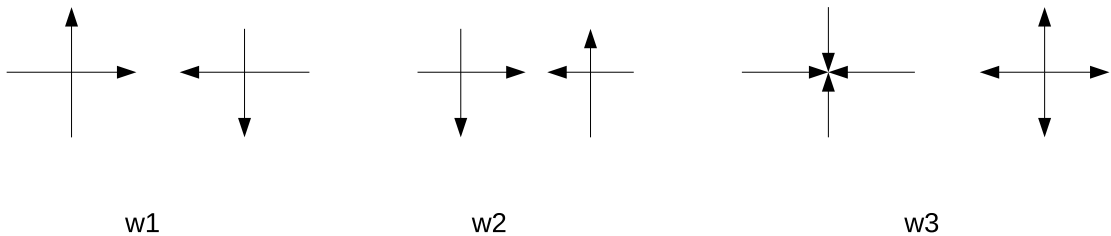}}
 \caption{Boltzmann weights of the Six-vertex model}
\label{Fig-Six-Vertex}
\end{figure}

As we have seen in the previous Section, the one-dimensional fermionic model with nearest neighbors interaction
displays a Mott transition. This property can be established through 
the integrability of the (one-dimensional) $XXZ$ spin chain, which is equivalent to the linearity of the associated free boson system.
Our strategy for discussing the Mott transition in the corresponding two-dimensional spinless fermionic 
model on the square lattice (\ref{Model-Ferm-2d}) is to show that the property of integrability could also
be extended to encompass this case. The discussion of the Mott transition in this system could then
follow the line of reasoning of the previous Section.
For reasons that will be clear latter, we will first consider the Six-vertex model \cite{Baxter1}), and later show its relation to the lattice fermionic model.  
The Six-vertex model is a statistical model on a two-dimensional lattice, on which a classical electric current 
defined on each link can interact with other currents at the lattice sites. Each site of the lattice 
(which we will refer to as a vertex) may be in one of the six possible configurations shown in figure (\ref{Fig-Six-Vertex}).
The energy $\epsilon_v$ associated with a given vertex depends on the four current states at the edges only.
If we further impose a parity ($Z_2$) invariance, we are left with three possible vertices:
\bea
&& a=w_1=w_2=e^{-\beta \epsilon_1} \\
&& b=w_3=w_4=e^{-\beta \epsilon_2} \\
&& c=w_5=w_6=e^{-\beta \epsilon_3} \eea
Following Refs \cite{Sierra-Book} \cite{Baxter1}, one defines a vector space $V_i$ for each vertical link, another one 
in a given horizontal row $V _a$  (the so-called {\it auxiliary space}) and a {\it vertex operator} ${\bf R}$ such 
that its matrix elements are interpreted as the Boltzmann weights of a given vertex according to:
\beq
\langle \mu_{i+1},\a_i|{\bf R}| \mu_i ,\beta_i\rangle\ =\ W(\a_i\b_i,\mu_{i+1} \mu_i )\ , 
\eeq
where ${| a_i\rangle ,| b _i \rangle }$ denote the states on the vertical links, and ${| \mu_{i+1} \rangle, |\mu_i \rangle }$ the ones on the horizontal links. In this basis the R-Matrix of the Six-vertex model reads:
\bea  {\bf {R}}= \left [ \begin{array}{cccc}
a & 0   &  0       & 0 \\
0 & b  &  c & 0 \\
0 & c  & b  & 0  \\
0 & 0 & 0   &  a\  \label{R-matrix}  \end{array}  \right]\ . \eea
Following Baxter, we introduce the row-to-row transfer matrix which plays the role of a discrete-time evolution operator 
(the time variable is taken as flowing upwards from one row to the next). More precisely, the transfer matrix is an endomorphism: 
\beq
T:H_N \equiv V_1\otimes V_2\otimes .......\otimes V_N \flecha  V_1\otimes V_2\otimes .......\otimes V_N\ ,
\eeq
defined by:
\beq
T\ =\ {\rm Tr}_a[{ \bf R}_{aN} {\bf R}_{a (N-1)}.....{\bf R }_{a1}]\ , 
\eeq
where the trace ${\rm Tr}_a$ is taken on the auxiliary space. The integrability of the model is guaranteed by the existence of a set of mutual commuting row-to-row transfer matrices. In fact, it has been shown  \cite{Baxter1} that for two sets of Boltzmann weights, $(a,b,c)$ and  $(a', b',c')$ the transfer matrices of the Six-vertex model satisfy:
\beq
[\ T(a,b,c)\ ,\ T(a',b',c')\ ]\ =\ 0 \Leftrightarrow \Delta_{6v}\ =\ \Delta '_{6v}\ , 
\eeq
where $\Delta_{6v}=(a^2+b^2-c^2)/(2ab)$ ($\Delta'_{6v}$ is defined in an analogous way) is the so-called {\it anisotropy parameter}. 
This condition is equivalent to the existence of solutions of the Yang-Baxter equation  \cite{Baxter1}:
\beq
{\bf R}_{12}(u){\bf R}_{13}(u+v ){\bf R}_{23}(v)\ =\ {\bf R}_{23}(v){\bf R}_{13}(u+v){\bf R}_{12}(u)\ , 
\label{Yang-Baxter} \eeq     
where $u$ is the spectral (uniformization) parameter, and each operator ${\bf R}_{ij}$ acts non-trivially on 
$V_i\otimes V_j$. This equation can also be written in the form of the commutation relations for the quantum $L$-operators
\beq
{\bf R}_{12}(u-v){\bf L}_{13}(u){\bf L}_{23}(v)\ =\ {\bf L}_{23}(u){\bf L}(v)_{13}{\bf R}_{12}(u-v)\ . 
\label{LLR-RLL}\eeq  
As it is known \cite{Sierra-Book}, the transfer matrix of the Six-vertex model is related to the Hamiltonian of the $XXZ$ model by:
\beq
T(\mu)=e^{-\mu H_{xxz}(\Delta)}\ , 
\eeq
which also shows that the Yang-Baxter equation implies the integrability of the $XXZ$ spin model.

Moreover, equation (\ref{Yang-Baxter}) may be considered as an equation among operators, whose solutions define integrable planar lattice models, {\it i.e.}, $(2+0)$ dimensional statistical systems or $(1+1)$ dimensional quantum systems. 
Within this approach, the Six-vertex model may be considered as one specific solution of the Yang-Baxter equation when the space $V_i$ is the representation space of spin $1/2$ particles. Although a classification of the solutions of (\ref {Yang-Baxter}) is not known, some solutions have been found. These solutions are related to the quantum deformations of Lie algebras or, more precisely to the quantum deformations of the universal enveloping Lie Algebras also called {\it quantum groups}. In fact, the universal Yang-Baxter equation :
\beq
{\bf R}_{12}{\bf R}_{13}{\bf R}_{23}\ =\ {\bf R}_{23}{\bf R}_{13}{\bf R}_{12} \ ,
\label{Yang-Baxter-Universal} 
\eeq
which is the Yang-Baxter equation independent of the spectral parameter, has the so-called universal $R$-matrices as solutions. For example, when $V_i$ is the space representation of spin $1/2$, Drinfeld  \cite{Drinfeld} has 
given a solution:
\beq
R\ =\ q^{H\otimes H/2}\sum_{n=0}^\infty \frac{(1-q^{-2})^n}{[n]_q!} 
q^{\frac{n(1-n)}{2}} q^{n H/2} (X_+)^n\otimes q^{-n H/2}(X_-)^n\ , 
\eeq
where $q$ is a complex parameter, $ [n]_q= (q^n-q^{-n})/(q-q^{-1})$ is a $q$-number 
and the generators $X_+$, $X_-$ and $H$ satisfy the commutation relations:
\bea
&& [X_+,X_-]=\frac{q^H-q^{-H}}{q-q^{-1}} \label{commutation-QG-1}\\
&& [H,X_+]=2X_+ \label{commutation-QG-2} \\
&& [H,X_-]=2X_- \ , \label{commutation-QG-3}
\eea
with co-products:
\bea
&& \Delta(X_\pm)=X_\pm \otimes q^{H/2} + q^{-H/2} \otimes X_\pm \\
&& \Delta(H)=H\otimes 1+1\otimes H \, 
\eea
which define the quantum group $U_q(sl(2))$ (note that the co-multiplication operator $\Delta$ 
should not be confused with the anisotropy parameter, as is clear from the context). 
Furthermore, we define the operators \cite{Sierra-Book} :
\bea
&& E_1=e^\mu S^+ \quad F_1=e^{-\mu}S^- \quad H_1=2S^z \\
&& E_o=e^\mu S^-  \quad F_0=e^{-\mu}S^+ \quad H_o=-2S^z\  ,\eea
where , $S^\pm$ are the raising (lowering) operators of the spin-$1/2$ particle and $x=e^{\mu}$ is the affinization parameter. These operators define an irreducible representation ($e^{\mu},1/2$) of the affine algebra $\widehat{sl(2)}$. 
In this context, the $R$-Matrix act as an {\it intertwiner} between the tensor product of two representations : 
\beq
R(e^{\mu_1},e^{\mu_2}) \Delta(g)=\Delta'(g) R(e^{\mu_1},e^{\mu_2})\ ,
\eeq
where  $g$ is any element of the Quantum Group and $\Delta'$ is the inverse co-product, {\it i.e.}, the co-product 
composed with the operator permuting vector spaces. This $R$-matrix has the form:
\bea 
R(e^{\mu_1},e^{\mu_2})= \left [ \begin{array}{cccc}
qx-q^{-1}x^{-1} & 0   &  0       & 0 \\
0 & x-x^{-1}  & q-q^{-1} & 0 \\
0 &  q-q^{-1}  &  x-x^{-1}     & 0  \\
0 &  0    & 0         &qx-q^{-1}x^{-1} \  \label{R-matrix-6v-2}  \end{array}  \right]\ , \eea 
where $x=e^{\mu}$, ${\mu=\mu_1-\mu_2}$. This $R$-matrix coincides with that of the Six-vertex model for the parametrization $a=\sinh(u+i\gamma)$ ,$b=\sinh(u)$ $c=i\sin\gamma$ $q=\exp{(i\gamma)}$ (for details see \cite{Sierra-Book}), so that the Six-vertex model possesses symmetry $\Uqa$. Moreover, it have been shown in \cite{Davis-Jimbo} that the Hamiltonian of the $XXZ$ model in the thermodynamic limit commutes with the Affine Quantum Group $\Uqa$ and that the space of states is identified with the tensor product of level $1$ highest and level $(-1)$ lowest representations of $\Uqa$. Besides, the 
corresponding $L$-operator is a $q$-deformation of the fundamental $L$-operator of the $XXX$ (Heisenberg) spin chain, given by \cite{Faddeev}:
\bea L^{xxz}_{n,a}= \left [ \begin{array}{cccc}
xq^{S^z _n}-x^{-1}q^{-S^z _n} & (q-q^{-1})S^-_n   &     \\
(q-q^{-1})S^+_n &      xq^{-S_z}-x^{-1}q^{S^z _n}    \  \label{Lxxz}  \end{array}  \right]\ . 
\eea 
This Lax operator, together with the $R$-matrix (\ref{R-matrix-6v-2}) satisfies the $LLR=RLL$ condition (\ref{LLR-RLL})). 
It is possible to rewrite this condition introducing
\beq
\tilde{L}(x)=Q(x)L(x)Q^-{1}(x) \qquad \tilde{R}=Q(x)Q(y)R(x/y)Q^{-1}(x)Q^{-1}(y)\ ,
\eeq
where 
\bea Q(x)= \left [ \begin{array}{cc}
x^{1/2} & 0       \\
0 &      x^{-1/2}   \  \label{matrix-Q}  \end{array}  \right]\ , \eea 
which yields:
\bea
 \tilde{R}= \left [ \begin{array}{cccc}
qx-q^{-1}x^{-1} & 0   &  0       & 0 \\
0 & x-x^{-1}  & x^{-1}(q-q^{-1}) & 0 \\
0 &  x(q-q^{-1})  &  x-x^{-1}     & 0  \\
0 &  0    & 0         &qx-q^{-1}x^{-1} \  
\label{R-matrix-Uqa}  \end{array}  \right]\ \ .  \eea
\subsection { Three-dimensional structure of Quantum Groups and Vertex Models}

\begin{figure}[ht!]
\centering
\resizebox{7.cm}{!}{ 
\includegraphics[clip]{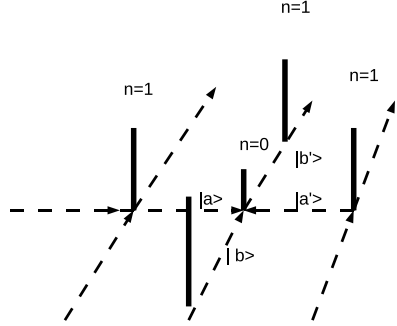}}
 \caption{\small{Graphical representation of the sates  $|n=0\rangle$ and $|n=1\rangle$ belonging to the Fock spaces at each lattice site. the arrows correspond to the states $|a_i\rangle$ , $|a'_i\rangle$ ,$|b_j\rangle$ ,$|b'_j\rangle$   }}
\label{figure-Fock}
\end{figure}

The $(2+1)$-dimensional analogue of the Six-vertex model, is a  `quantum-vertex model' where the classical weights $e^{-\beta \epsi _a}$, $e^{-\beta \epsi _b}$ $e^{-\beta \epsi _c}$, should be replaced by `quantum vertex operators' defined on the lattice Hilbert space. In order to define this vertex model, we need first to consider a quantum lattice \cite{Sergeev-Quantum-Lattice} defined as follows: for each lattice site $(x_i,y_j)$,  we define a Fock space ${\it  F}_{ij}$ and a set of  representation spaces of spin $1/2$ particles $V_i$, $V_j$, $V_{i+1}$ and $V_{j+1}$ on each link joining two lattice sites. The states of the link in the lattice are arrows (as in the Six-vertex model), and the states in the Fock space ${\it  F}_{ij}$ are labeled by the number of particles in the site $|n_{ij}\rangle$ as shown in figure \ref{figure-Fock}. Then, we assign a scattering amplitude (and a vertex operator) to each lattice site by: 
\beq
S_{\a_{i},\b _j} ^{\a_i',\b _j'}\ =\ \langle \a'_i \b '_j n'_{ij} |{\bf L}_{V_i ,V_j, F_{ij}}| \a_i ,\b_j n_{ij} \rangle =L_{i,j,n} ^{i' j',n'}\label{Boltzman-model}
\eeq
where ${\bf L}_{V_i ,V_j, F_{ij}}$ is a `three dimensional Lax operator' acting  on the spaces $V_i\otimes V_j\otimes F$. 
Furthermore, it is possible to define layer-to-layer transfer matrices $T_{m n}(\{\l\},\{\mu\})$, where the pair $(m,n)$ labels the rows and columns of a given layer and $(\{\l \},\{\mu \})$ are the spectral parameters, associated to the rows and columns, respectively, by:
\beq
T_{mn}\ =\ {\rm Tr}_{V_x\otimes V_y}[\prod_i \prod_j {\bf L}_{V_i,V_j,F,}(\l_i,\mu_j)]=T_{mn}(\{\l \},\{\mu\})\ , 
\label{layer-to-layer}\eeq
where $V_x=\otimes_i^n V_i$ and $V_y=\otimes_j^m V_j$,$\{\l \}=\{\l_1,\l_2,....\l_n\}\{\mu\}=\{\mu_1,...,\mu_m\}$.
Here, the layer-to-layer transfer matrix plays the role of a temporal evolution operator in an unitary time step, where the temporal axis coincides with the direction perpendicular to the layer.
For  quantum systems in $(2+1)$-dimensions (and also for three-dimensional statistical systems), the integrability is  guaranteed by the commutativity of the layer-to-layer transfer matrices. The integrability is also encoded in the  the so-called  Zamolodchikov Tetrahedron Equation (TE), which is the three-dimensional analogue of the Yang-Baxter equation, or equivalently in terms of the Lax operators, through the three-dimensional analogue of the $LLR-RLL$ equation:
\bea
&& {\bf R}_{abc} {\bf R}_{ade}{\bf R}_{bdf}{\bf R}_{cef}={\bf R}_{cef}{\bf R}_{bdf}{\bf R}_{adc}{\bf R}_{abc} \ ,\label{tetraeq}\\
&& {\bf L}_{12,a} {\bf L}_{13,b}{\bf L}_{23,c}{\bf R}_{abc}= {\bf R}_{abc}{\bf L}_{23c}{\bf L}_{13,c}{\bf L}_{13,b} \label{LLLR-RLLL}\ ,\eea
where the operators $R_{ijk}$ define the mapping  
$ R_{abc}: F_a \otimes F_b \otimes F_c \rightarrow F_a \otimes F_b \otimes F_c $, and the operators $L$ act on $V=V_1\otimes V_2\otimes V_3 \otimes F_a\otimes F_b\otimes F_c$.  
If the Fock space $F_a$ is considered as the representation space of some algebra ${\cal A}$, then the operators $L$ can be represented as operator-valued matrices acting on $V_i\otimes V_j$ such that their coefficients are given in terms of the generators ${v_a}$ of the algebra ${\cal A}$ and some complex parameters $s_a$, so that the equation (\ref{LLLR-RLLL}) takes the form of a local-Yang-Baxter-equation:   
\beq
 L_{12}({\bf v}_a,s_a) L_{13}({\bf v} _b,s_b) L_{23}({\bf v} _c,s_c)= L_{23}({\bf v'}_c,s_c)  L_{13}({\bf v'}_b,s_b)  L_{1a}({\bf v'}_a,s_a)\ .  \label{Local-Yang-Baxter}
\eeq
For classical systems, the algebra ${\cal A}$ is chosen to be the Poisson Algebra $P$. In this case, the matrix $L$ 
which solves the local Yang-Baxter equation (\ref{Local-Yang-Baxter}) is given by
\bea 
L_{1,2}(k_a,a_a,a^*_a)= \left [ \begin{array}{cccc}
1 & 0   &  0       & 0 \\
0 & k_a  &  a_a^{*} & 0 \\
0 & -a_a  &  k_a     & 0  \\
0 &  0    & 0         & 1\  \label{lunif}  \end{array}  \right] \ ,
\eea
where the indices in the second space enumerate the two-dimensional blocks , while those for the first space enumerate the elements inside the blocks and $k=1-a^*a$ . Moreover,
\beq
\{a^*_i,a_j\}_{PB}=2 \delta_{ij} \qquad \{k_i,a_j\}_{PB}=\delta_{ij}k_i a_j \qquad  \{k_i,a^*_j\}_{PB}=-\delta_{ij}k_i a^*_j \eeq
where $\{\ ,\ \}_{PB}$ denote the Poisson brackets. It was shown in \cite{Bazhanov-1} that equation (\ref{Local-Yang-Baxter}) defines a canonical transformation (automorphism) of the triple tensor product of the Poisson algebra. This solution correspond to the classical three-wave problem, {\it i.e.}, the linear propagation of three-dimensional waves along to three mutually perpendicular axes.  

In quantum systems, we expect that the Algebra ${\cal A}$ will be the either bosonic or fermionic. This is actually the case for free Boson or Fermion systems. 
However, we are interested in {\it interacting solutions} of the $TE$. Recently, a new solution to the $TE$ associated with the three-dimensional structure of the  affine quantum group $\widehat{U_q(sl(n))}$ has been found in \cite{Bazhanov-1} \cite{Bazhanov-2}. The new solution  may be understood as the quantization of either the classical three-dimensional wave problem or the quantization of fluctuations of extended spatial objects. It amounts to taking $L$-operators as block matrices with two-dimensional blocks, in which matrix indices in the second space enumerate the blocks while those for the first space enumerate the elements inside the blocks.  
\bea L_{i,j}({\cal A}_v) =  \left [ \begin{array}{cccc}
1 & 0 & 0          & 0 \\
0 & \l_v  k_v & a_v^{\dagger} & 0 \\
0 & -q^{-1}\l_v \mu a_v  &    \mu _v k_v     & 0  \\
0 & 0 & 0          & -q^{-1}\l_v \mu_v\ \ , \label{L-uniformizado}  \end{array}  \right] \ .
\eea 
with $i,j=1,2,3 $ , $v=a,b,c$ and where now $(k,a^\dagger,a)$ are quantum operators acting on the Hilbert space and the algebra ${\cal A}$ is the $q$-oscillator algebra defined by \cite{Bazhanov-1}:
\bea
&& q a^\dagger a -q^{-1}a a^\dagger=q-q^{-1} \quad [h,a^\dagger]=a^\dagger \quad [h,a]=-a \\
&& k^2=(1-a^\dagger a) \qquad k=q^h \eea

\begin{figure}[ht!]  
\centering
\resizebox{15.cm}{!}{ 
\includegraphics[clip]{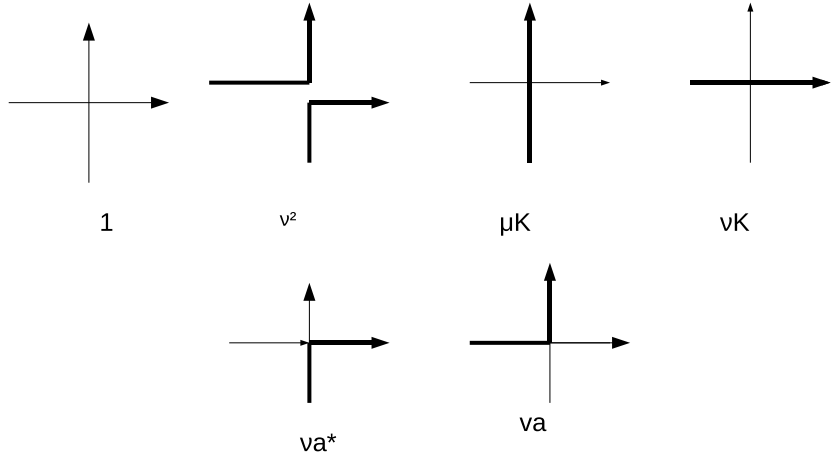}}
 \caption{Weights of the Quantum-Vertex-Model}
\label{Fig-Quantum-Vertex}
\end{figure}

The above solution allows one to define  a `quantum vertex model' by assigning `quantum vertex operators' $f_j$ at each vertex on the lattice, according the rules shown in the figure (\ref{Fig-Quantum-Vertex} ), where $\nu^2=-q^{-1}\l \mu$.
We now map the square lattice onto a torus, and let $C_a$ and $C_b$ be the two basic homotopy cycles on that torus, such  that each homology cycle corresponds to a path along one coordinate axis of the square lattice .  Any path on the torus belongs to a  given homotopy class $P \sim n C_a+ m C_b $, and it is possible to define \cite{Sergeev-Bose-Gas}: 
\beq
T_{n,m}=\sum_{P } \sum_{j \in P} f_j\ , \eeq 
where the sum over $P$ means the sum over different paths,  
which is exactly the equation (\ref{layer-to-layer}).
The commutativity of the layer-to-layer transfer matrices follows from another (related) Tetrahedron Equation \cite{Sergeev-integrability-q-oscilator} \cite{Bazhanov-1}:
\bea
&& M_{ii'}(\it{H}_0\mu/ \mu')M_{j j'}(\it{H}_0\l/ \l')L_{ij}(A_v,\l,\mu)L_{i'j'}(A_v,\l,\mu)=\nn \\
 && L_{i'j'}(A_v,\l,\mu)L_{ij}(A_v,\l,\mu)M_{j j'}(\it{H}_0\l/ \l')M_{ii'}(\it{H}_0\mu/ \mu')\ , \label{Tetrahedron-2}\eea
where the matrix elements of $M({\it H}_0,\zeta)$ belong to an additional copy of the $q$-oscillator algebra denoted 
by ${\it H}_0$ and :
\bea 
M_{i,j}({\it H}_0) =  \left [ \begin{array}{cccc}
\zeta ^{h_O} & 0 & 0          & 0 \\
0 & \l_0 (-q\zeta)^{h_0} & \nu_0 \zeta^{-1/2+h_0}a_0^{\dagger} & 0 \\
0 & \mu_0\zeta^{1/2+h_0} a_0  &    \mu _0 (-q\zeta)^{h_0}      & 0  \\
0 & 0 & 0          &  \mu_0\zeta^{h_0} \label{M-operator}  \end{array}  \right] \ ,
\eea
with $\nu_0 ^2=q^{-1}\l_0 \mu_0$. Equation (\ref{Tetrahedron-2}) can be verified directly from the operator (\ref{M-operator}), and the commutativity  of the layer-to-layer transfer matrices follows from its definition and the use of equation  (\ref{Tetrahedron-2}).  
In the cubic lattice with boundary conditions( where the third dimension corresponds to the temporal axis) , the solutions of the $TE$ given in (\ref{L-uniformizado}) \ref{M-operator} have the following properties:   
\begin{itemize}
\item For $q<1$, they yield the Fock space representation of the $q$-oscillator algebra:
\beq
a|0\rangle =0 \quad |n\rangle=\frac{(a^\dagger)^n}{\sqrt{(q^2;q^2)}}|0\rangle \quad  k|n\rangle=q^{n+1/2}|n\rangle \eeq
where $(x;q^2)=(1-x)(1-q^2x)......(1-q^2x)$. 

It is possible to construct  the elements of the $R$-matrix in the basis of the $q$-oscillator algebra 
$\langle n'_1,n'_2,n'_3|R|n_1,n_2,n_3\rangle $. The resulting $R$-matrix is non-degenerate in $F^{\otimes^3}$.

\item  The above solution implies that the standard Yang-Baxter equation is satisfied. One defines the operators
\bea
&& R_{bc}=Tr_{F_a}[R_{ab_1,c_1 }R_{ab_2,c_2 }.....R_{ab_n,c_n }]\label{Def-Rbc} \\
&& L_{vb}=Tr_{V_0}[L_{0,1,b_1} L_{0,2,b_2}.....L_{0,n,b_n} ]\label{Def-Lvb}\eea
where $F_b=F_{b_1}\otimes F_{b_2}\otimes....F_{b_n}$,  $F_c=F_{c_1}\otimes F_{b_2}\otimes....F_{c_n}$, which involves the product of $n$ operators along the `third direction, {\it i.e.}, the direction labeled by  `a'. 
Due to the fact that the $R$-matrix is non-degenerate in $F_a\otimes F_b\otimes F_c$ , the $TE$ implies that 
these operators satisfy  
\bea
R_{bc}R_{bd}R_{cd} &=& R_{cd}R_{bd}R_{bc} \label{Yang-Baxter-projection}\\
L_{Vb}L_{Vc}R_{bc} &=& R_{bc}L_{Vc}L_{vb} \label{LRR-RLL-projection} \ .\eea 
This construction yields a {\it projection} from the $TE$ to the $YB$ ones.
Using a similar projection, the equation \ref{M-operator} implies that there also exists another Yang-Baxter equation (for details see \cite{Bazhanov-1}):
\bea
R_{V_i,V_j}L_{V_i,b}L_{V_j,b}=L_{V_i,b}L_{V_j,b}R_{V_i,V_j}\label{Yang-BaxSix-Vertex-2}\ . \eea

\item   The new solution is associated with the affine quantum group $U_q(\widehat{sl(n)})$, where $n$ is the range  of the third dimension, {\it i.e}, $n$ is the number of two-dimensional layers.

\end{itemize}
For simplicity we now take $\l=1$ $\mu=1$  and the indices $1,2, 3$ denoting the quantum spaces $a,b,c$. 
Inserting the operator $L$ (\ref{L-uniformizado}) in the local Yang-Baxter equation (\ref{Local-Yang-Baxter}), we obtain 
the explicit mapping $R_{123}$  given by:
\bea
&& k'_2(a^\dagger_1)'=k_3a^\dagger_1- k_1a^\dagger_2a_3 \qquad  k'_2a'_1=k_3a_1- k_1a_2a^\dagger_3 \nonumber\\    
&&(a^\dagger_2)'=a^\dagger_2a^\dagger_3 + k_1k_3a^\dagger_2 \quad \qquad a_2'=a_2a_3 + k_1k_3a_2 \label{Map-R123} \\ 
&& k'_2(a^\dagger_3)'=k_1a^\dagger_3-  k_3a_1a^\dagger_2     \qquad   k'_2a'_3=k_1a_3- k_3a^\dagger_1a_2 \ .\nonumber     
\eea
Equations (\ref{Map-R123}) are the Heisenberg equations of motion for the quantum operators $(a_i,a^\dagger_i,k_i)$, where we have denoted with primes those  operators that evolve forward in time $t'=t+\Delta t$ which describe the time evolution of the quantum three-wave problem.

\subsection{Integrability and the two-dimensional Mott transition} \label{Integrability-2D-mott}

In this Section we will come back to the two-dimensional Fermion model defined by (\ref{Model-Ferm-2d}) to study its symmetries and integrability. We first recall (Section (\ref{section-1})) that it is equivalent to a two-dimensional $XXZ$ spin model:
\beq
H_F=H_{XXZ}=-\sum_{ij} [\  S^x_i S^x_J +   S^y_i S^y_J-\Delta S^z_i S^z_j\ ]\ . \nonumber \eeq
On the other hand, the quantum vertex model defined in the previous Section has the remarkable property that can be projected from three to two dimensions. This means that the equations:
\bea
 L_{Vb}L_{Vc}R_{bc} &=& R_{bc}L_{Vc}L_{Vb}  \\ 
L_{V_i,b}L_{V_j,b}R_{V_i,V_j}&=&R_{V_i,V_j}L_{V_i,b}L_{V_j,b}\ , \eea
can be interpreted as arising from a two-dimensional system. Moreover, two remarkable properties  of the quantum vertex model and the associated three-dimensional structure of quantum groups have been discussed in \cite{Bazhanov-1}:
\bea
&&L_{Vb}=\otimes_{i=1}^n {\cal L} (\omega_k, \lambda ,\{\mu_i\}) \\   
&& T_m=Tr_{\pi_{\omega_k}}[{\cal L} (\omega_k ,\l_m \{\mu_i\})\, ..... \quad {\cal L}(\omega_k,\l_1, \{\mu_i\})]. \eea
The first equation shows that the operator $L_{Vb}$ decomposes into a direct sum of the fundamental ${\cal L}$-operators  ${\cal L}^{sl(n)}$ of the  affine quantum group $U_q(\widehat{sl(n)})$, where $\omega_k$ is the highest weigh of the representation $\pi_{\omega_k}$ .
The second one, shows that the row-to-row transfer matrix of the quantum vertex model can be reconstructed from the fundamental $\cal{L}$-operators.

The Quantum vertex model is stationary for $L_{12}(a,a^\dagger,k)=L_{12}(a',{a^\dagger }',k')$, {\it i.e.}, for the case when the patterns for two consecutive time slices are identical so that the third dimension has range $n=2$. In this case, the operator $L_{vb}$ is:
\bea
L_{Vb}(u) = \left[  \begin{array}{ccc}
1+u\l_1 \l_2 q^{h_1+h_2}  & 0 & 0   \\
0                  & {\cal L} (\frac{1}{2},\mu) & 0 \\
0          &0           & \mu_1 \mu_2 (q^{h_1+h_2}+  q^{-2} u\l_i\l_2)\ \ , \label{Lvb}  \end{array}  \right]\eea
and 
\bea 
{\cal L} (\frac{1}{2},u) =  \left[\begin{array}{cc}
\mu_1(q^{h_1}- u\l_1 \l_2 q^{h_2 -1}) & -q^{-1}\l _1\mu_1 a_1 a^\dagger_2  \\
 -q^{-1} u \l_2 \mu_2 a^\dagger_1 a_2  &\mu_2(q^{h_2}-u\l_1 \l_2 q^{h_1 -1})  \  \label{L1/2}  \end{array}  \right] \ ,
\eea
For $h_1=h_2=1/2$ we have:
\bea
q^{h_1}=\left[ \begin{array}{cc}
 q & 0 \\
0 & 1  \end{array}  \right]\ , 
\,  
q^{h_2}=\left[ \begin{array}{cc}
 1 & 0 \\
0 & q  \end{array}  \right]\ , 
\,
 a_1 a^\dagger_2=\left[ \begin{array}{cc}
 0 & 0 \\
1-q^2 & 0\end{array}  \right]\ , 
\,
 a^\dagger_1a_2=\left[ \begin{array}{cc}
 0 & 1-q^2 \\
0 & 1\end{array}  \right] \nn 
\eea
If $\l_1=\l_2=1$ and $\mu_1=\mu_2=1$,  the last operator is the $R$-matrix (\ref{R-matrix-Uqa}) of the $XXZ$ spin chain( or Six-vertex $R$-matrix) . Since is possible to project the quantum vertex model onto any lattice direction, any row or column should give rise to a $XXZ$ spin chain, and one can expect that this system will be equivalent to a two dimensional $XXZ$ model.  (esta propiedad es algo asi como una generalizacion de la propiedad de invarianza modular) 

\vspace{0.5 cm}

We are now ready to map among each other the original two-dimensional fermion model, the quantum vertex model and the two-dimensional $XXZ$ spin model. The two-dimensional lattice fermion model describes charge density waves propagating on the lattice of the underlying electrons above ( and below) the half-filling state. 
Viewed at a fixed time, the wave vectors of these charge density waves cross themselves at each lattice site, defining a `Six-vertex model'. The half-filling condition means that the Fock space of the $q$-oscillator algebra must be restricted to: $\{|n\rangle ={|1\rangle, |-1\rangle} \}$. These states are in one-to-one correspondence with the states $|\uparrow \rangle$ $|\downarrow\rangle$ of the two-dimensional spin model.  Therefore, the quantum ( Fock) space is also in correspondence with the representation space of a spin $1/2$ particle. Introducing  this information in the equations (\ref{Yang-BaxSix-Vertex-2}) (\ref{Lvb}) (\ref{L1/2}), we see that for each line on the lattice we can define an $XXZ$ (one dimensional) spin chain corresponding to the wave propagation of the CDW along this line. This fact allows us to identify the parameter $\Delta=U/t$ (representing the normalized Coulomb interaction) of the one-dimensional Fermion model 
with $\Delta=-(q+q^{-1})/2$, where $q=\exp{(i\gamma)} $ is the deformation parameter of the quantum group.

Moreover, let us consider an interaction star , {\it i.e.}, the interaction among a central spin, labeled by the index $\a$ an its nearest neighbors, labeled by the index $\b$. A direct calculation  shows that:
\bea
\frac{i}{2}\frac{d}{d u} \ln[1/a(u)Tr_{\a,\b}{\cal L}(u)_{1/2} ({\bf r}_\a){\cal L}(u)_{1/2}({\bf r}_\b) ]=-H_{\a,\b} \ , \eea 
where $a(u)=\sinh(u+i\gamma)$, and we have taken the parameters $u=v=1$. $Tr_{\a,\b}$ is the trace over the space $V_{i\a}\otimes V_{j\a}\otimes V_{i\b}\otimes V_{j\b}$ and $H_ {a\,\b}$ is the element $(\a,\b)$ of the two-dimensional $XXZ$ Hamiltonian. Now taking :
\bea
L_{ij}(\l_i,\mu_j)=     
\begin{cases} 
1 &\textit{if $|r_\b-r_\b|>1$}\\ 
L(a,a^\dagger,k,{\bf r}_\a)  &\textit{${\bf r}_\a,{\bf r}_\b$  are first neighbors}
\end{cases}
\eea
we have that the partition function of the quantum vertex model (QVM) is:
\beq
Z(QVR)=Tr_F^{nm} \prod_{\a \b}[ Tr_{V_\a\otimes V_\b}L(a_\a,a^\dagger_\a,k_\a, {\bf r}_\a)  L(a_\b,a^\dagger_\b,k, {\bf r}_\b)] \ ,  
\eeq 
where, $F^{nm}= F^{\otimes nn}$ is the complete Fock space of the square lattice with $n$ rows and $m$ columns. Tracing 
over the vector  space $V_0$ (there is nothing special about the `line zero', so that we can also trace over any row space$V_1$,$V_2$....$V_n$) and using equation (\ref{Def-Lvb}), we have:
\bea
&& Z(QVM)=Tr_{F^{nm}} \prod_{\b}[Tr_{V_\b}(L_{vb}({\bf y}_\b)L_{vb} ({\bf y}_{\b+1})^N)] \\ 
&&       =Tr_{F^m} \prod_{\b} [Tr_{V_\b}( R^{xxz}({\bf y}_\b)R^{xxz} ({\bf y}_{\b+1}))^N]\\
&&      =Tr_{F^m} [\prod_{\b}e^{\sum_\a H_{\a,\b}}] \\ 
&&      =Tr_{F} [e^{\sum_{\a , \b} H_{\a,\b}}]\\
&&       =Z(H^{2-D}_{XXZ} ) \eea
where in the second line we have taken into account the fact that we are using the two-dimensional (spin $1/2$) representation of $\uqa$. We arrive at the interesting identity:
\beq
Z(QVM)=Z(H_{xxz}^{2d})\qquad , \qquad H^{2d}_{xxz}=H_F\ . \label{equivalencia}\eeq 
This equation (\ref{equivalencia}) shows that the two-dimensional $XXZ$ spin lattice and the Lattice fermion system (\ref{Model-Ferm-2d}) have Quantum Group Symmetry $\uqa$. 

\begin{figure}[ht!]
\centering
\resizebox{10.cm}{!}{ 
\includegraphics[clip]{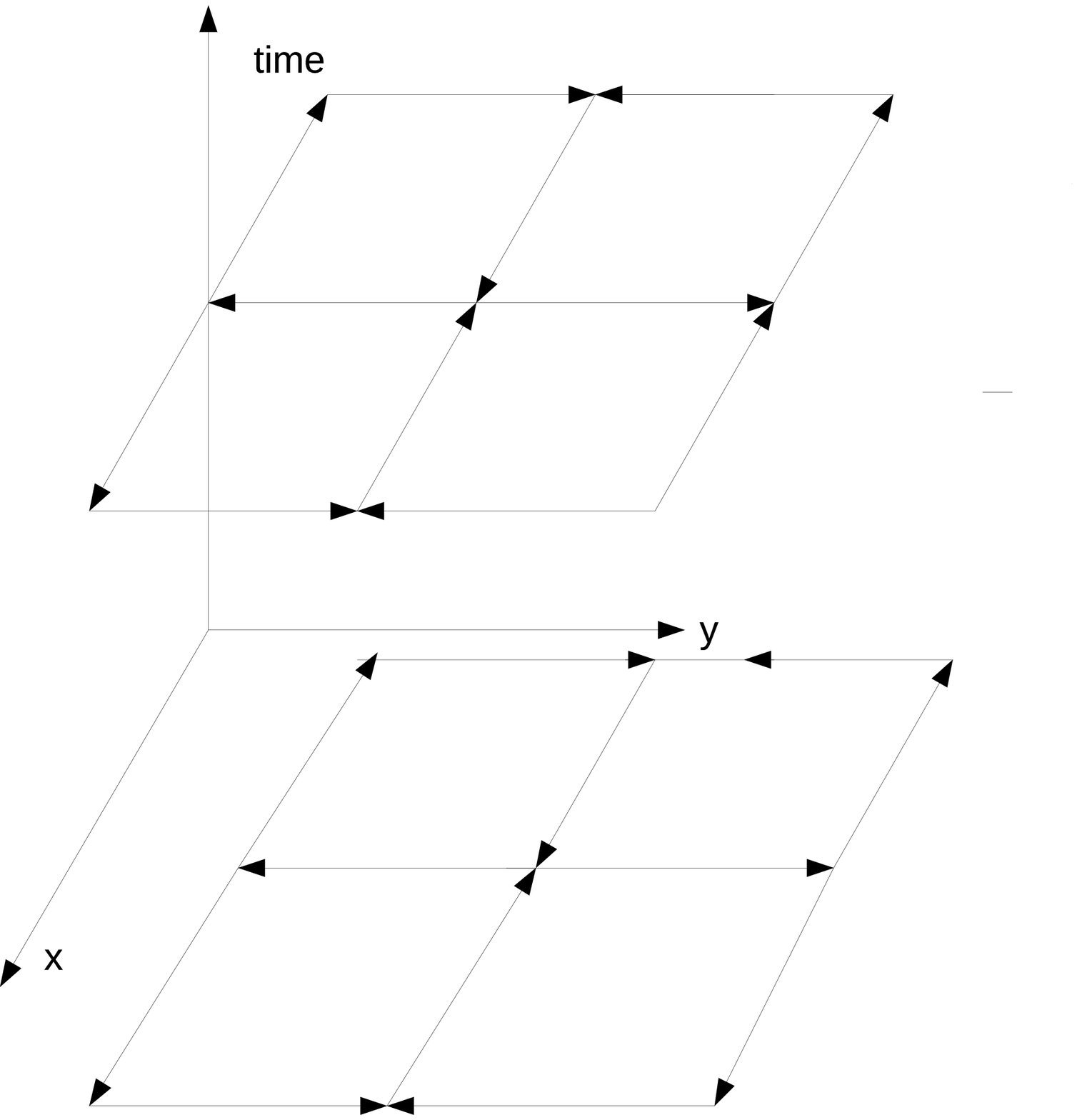}}
 \caption{\small{Current patterns of the Fermion model at two different times.}}
\label{Two-Layer}
\end{figure}

Summing up: the two-dimensional lattice Fermion model, viewed at two different times, can be considered as a two-layered 
three-dimensional system where the third direction coincides with the temporal axis (see figure \ref{Two-Layer}). From an intuitive point of view, one can assign an arrow to the time propagation direction of the charge density waves of the underlying fermions. At the stationary point, where the pattern of arrows does not change with time, these arrows define a vertex model with quantum group symmetry $\widehat {U_q(sl(2))}$. Alternatively, one can say that due the projection-like property (\ref{Yang-Baxter-projection}), the two-dimensional $XXZ$ spin system may be decomposed in a consistent way into two one-dimensional chains, so that the entire system will have a Quantum Group symmetry ($(\uqa)^{\otimes N}$). 
The property that $(2+1)$-dimensional  quantum systems in square lattice with periodic boundary conditions reduce to $(1+1)$-dimensional quantum chains, implying that the ($(\uqa)^{\otimes N}$) symmetry reduces to the $\uqa$ Quantum Group symmetry, was first noted in \cite{Bazhanov-1} \cite{Sergeev-Bose-Gas}.      


\section{Construction of the Effective Field Theory} 

As we have shown in the previous Section, the two-dimensional fermionic system defined by (\ref {Model-Ferm-2d}) has a $(2+1)$ Quantum-Group symmetry $ U_q(\widehat{sl(2)})$. 
Discrete time evolution is given by the transfer matrix $T_{mn}(\{\l_n\},\{\mu_m\})$. We have presented an 
explicit form for it in (\ref{layer-to-layer}) which is a solution of the Zamolodchikov Tetrahedron 
Equation, rendering the model integrable in $(2+1)$ dimensions. 
However, this solution is not completely transparent from the physical point of view.
To further study the system and this solution, we will follow the path of writing down an Effective Field Theory (EFT) formulation for it (for a review, see \cite{Polchinski}), valid  for long-distance and low-energy domains. EFTs provide useful frameworks for analyzing the behavior of many-body systems, specially near to a phase transition point, were universal properties are vastly dominating the physical properties of the systems under study. The general scheme for applying the method of EFTs starts by identifying the effective degrees of freedom that dominate the low-energy regime of a given system  and their characteristic symmetries. Note that these degrees of freedom are usually chosen by phenomenological reasons and could bear no resemblance with the microscopic degrees of freedom of any underlying model describing the system. One then proceeds to write down the most general local action in terms of second-quantized fields representing the selected
degrees of freedom which are consistent with the noted symmetries. 
In the case at hand, the construction of the EFT is obtained by choosing fermionic fields as the 
degrees of freedom. By using the vertex model representation of the lattice fermion model (\ref{Model-Ferm-2d}),and its connection to the Chern-Simons theory (and analytical continuation in the Chern-Simons coupling constant given in \cite{Witten-Vertex-Models}), we will now show that the corresponding Effective Field Theory is a double Chern-Simons theory with Gauge Group $U(1)$ and Quantum Group symmetry $\widehat {U_q(sl(2))}\otimes \widehat {U_q(sl(2))}$, which is the symmetry of the exact solution (\ref{layer-to-layer}) (\ref{L-uniformizado}). 

\subsection{The one-dimensional case}

For the sake of completeness and clarity of exposition, let us start by considering the EFT describing the one-dimensional Mott transition. As discussed in Section( \ref{section-1}), the one dimensional counterpart of the lattice fermion system (\ref{Model-Ferm-2d}) can be bosonized. Its effective degrees of freedom are the charge density waves of the underlying strongly correlated electrons, and for a general coupling ($\Delta$) the system has a $\widehat{u(1)}$ Kac-Moody symmetry \cite{Voit}. The Mott transition point ($\Delta=1$) coincides with the Lutter-Emery point, where the degrees of freedom are two chiral non-interacting fermions. At this point, the  equation (\ref{Lutter-Emery-ponit}) can be written in coordinate space on a circle domain as:
\beq
H=\frac{v}{2}\int_0 ^{2\pi R}dx [\psi_r^\dagger(x)(-i\de_x) \psi_r(x) +cc  + \psi_l^\dagger(x)(-i\de_x) \psi_l(x) +cc]\ ,   \eeq  
where $ \psi_r(x)$ ( $\psi_l(x)$) are the right (left) relativistic chiral fermions (Weyl fermions) moving on a circle 
of radius $R$ with speed $v$. Each chiral branch can be bosonized independently. Let us focus on one chiral component,say the right one. This can system has a quadratic action on a circle (of radius $R=1$) \cite{Floreanini-Jackiw}:
\beq
S=-\frac{k}{4 \pi} \int_{\infty}^{\infty} dt \int_0^{2\pi } (\de _t +v \de_x) \phi \de_x \phi\ , 
\eeq
where $k$ is the coupling constant, $\phi (x,t)$ is a real scalar field and $x$ denotes the coordinate along the circle of length $2\pi$ and $v$ is the `speed of light'. 
By coupling this one-dimensional chiral system to an external electromagnetic field $E$, the equation of motion of the bosonic field changes to \cite{Zemba-Dunne}:
\beq
\de_t Q=\frac{e}{k}E ,\label{Anomaly}
\eeq
where $E=\de_ t A_x- \de_x A_t$ is the electric field pointing along the circle and $A$ is the $U(1)$ electromagnetic potential. Equation (\ref{Anomaly}) displays the chiral anomaly of the Weyl Fermion. 
This description can be also obtained from the Abelian Chern-Simons theory with Lagrangian
\beq
L_{cs}\ =\ \frac{k}{4\pi} \epsilon ^{\mu \nu \rho} A_\mu \de_\nu A_\rho 
\eeq
defined on the disc $D$, whose boundary is the circle of radius $R=1$. By making a Gauge transformation  
$\delta A_\mu=e \de_\mu \lambda$, the action undergoes a variation:
\beq
\delta S_{CS}\ =\ \frac{ke} {4 \pi} \int \l(\de_t A_{\theta}-\de_\theta A_t) \ , 
\eeq
showing that $S_{CS}$ is not Gauge invariant on the boundary. Equivalently, there is a chiral current
\beq
J^r=\frac{\delta S_{cs}}{\delta A_r}=\frac{k e}{2\pi}E_\theta\ ,  
\eeq
that shows again the presence of the chiral anomaly of the Weyl Fermion for $k=1$ \cite{Zemba-Dunne}. 
The chiral anomaly of one chiral component is canceled by the corresponding anomaly of the
anti-chiral sector of the complete theory, which  is anomaly free. This implies that the Mott transition 
in a one-dimensional system is described by a double Chern-Simons theory :
\bea
S=\frac{k}{4\pi } \int_{D x R} d^3x  \epsilon ^{\mu \nu \rho}A^R_\mu \de_\nu A^R_\rho-\frac{k}{4\pi } \int_{D x R} d^3x  \epsilon ^{\mu \nu \rho}A^L_\mu \de_\nu A^L_\rho\ ,
\eea   
which contains two chiral Gauge fields ($A^R$, $A^L$ ) of opposite chiralities and where the value of the coupling constant is $k=1$.

\vspace{1 cm}
\subsection{Two-dimensional Effective Field Theory} \label{2D-EFT}
             
Now let us consider the EFT for the Mott transition in the two-dimensional square lattice. As it has been remarked before, one must identify the correct degrees of freedom dominating the low-energy regime and their characteristic symmetries. The two-dimensional Jordan-Wigner transformation shows that the lattice fermionic system can be also considered as a bosonic system. We choose these bosonic degrees of freedom as the characteristic ones for the low-energy regime of the system. 
Furthermore, these can be viewed in more physical terms as charge density waves, which we choose to construct the EFT. 
We look for an EFT with three-dimensional Quantum Group symmetry that is consistent with the properties established in  Section ( \ref{section-3}). Namely, that it should posses a kind of vertex model interpretation, that it should be parity-invariant and that it must be projection-able onto one-dimensional theories, each giving rise to a one-dimensional Mott transition. We recall that in these one-dimensional theories the degrees of freedom split naturally in two non-interacting chiral branches when the interaction parameter becomes $\Delta=1$. We will now show that this EFT is a lattice double Chern-Simon theory with $\uqa\otimes \uqa$ Quantum Group symmetry.        

\subsubsection{Vertex Models and Chern-Simon Theory}\label{Vertex-model-CS}
A direct connection between vertex models and Chern-Simons Gauge theory was first established in classical articles by Witten \cite{Witten-Vertex-Models} \cite{Witten-2} by using a non-Abelian Chern-Simons theory. Let us review this connection and consider a Gauge connection  $A=A_i t^a_i$ that belongs to an a Lie group $G$, where $t_i^a$ are the generators of the group. The non-Abelian Chern-Simons theory is defined by the action:
\beq
S_{CS}\ =\ \frac{k}{4 \pi } tr_R \int _{\cal M} [{\bf A} \wedge d{\bf A}\ +\ \frac{2}{3} {\bf A} \wedge {\bf A} \wedge  {\bf A} ]\ , 
\eeq
where ${\cal M}$ is an orientated topological manifold, and  $tr_R$ denotes the trace on the representation $R$ of the 
group $G$. The natural observables of this theory are Wilson loops: let us consider a link $L$ as a disjoint union of the circles $C_i$ and pick up a representation $R_i$ for each circle. The expectation value of the Link can be calculated as:
\beq
\langle L \rangle=\  \int DA e^{L_{cs}} \prod_i tr_{R_i} P e^{i \int A_i} \ .
\eeq
Here we can take ${\cal M}=\Sigma x R$, where $R$ represents the temporal axis and $\Sigma$ is a Riemann surface. In 
\cite{Witten-Vertex-Models} it has been shown that it is possible to define vertex models by replacing the classical currents living on the links of a lattice by Wilson lines in a Chern-Simons Gauge theory. Furthermore, it has been
shown that the expectation values of this Wilson lines can be calculated from the data in the CS theory. More specifically, taking the Gauge group $SU(2)$ and projecting the three-dimensional knots onto the plane, the `Boltzmann Weights' for these vertex models are given by:
\bea
&& W_{up}=q^{1/2} \delta_{s_1-s_3,0}-q^{-1/2}\delta_{s_1+s_2,0}q^{(s_2-s_3)} \\
&& W_{und}=q^{-1/2} \delta_{s_1-s_3,0}-q^{1/2}\delta_{s_1+s_2,0}q^{(s_2-s_3)} \\
&& W_{pc}=\epsilon_{s_1,s_2} q^{-s_1/2}\ , \label{Pesos-CS}\eea
where $ W_{up}$ ($ W_{und}$) denotes the Boltzmann Weight for the vertex in which the line labeled by the representations of $SU(2)$ denoted by $s_1$ and $s_2$, are above (below) the line labeled by $s_3$  and $s_4$. Furthermore, $W_{pc}$  denotes the Boltzmann weights for the pair-creation, and $q$ is the deformation parameter of the quantum group, which is related to the coupling constant of the Chern-Simons theory by $q=\exp{(i\pi/k)}$ . Note the a different factor in the definition of the deformation parameter of the Quantum Group with respect to \cite{Witten-Vertex-Models}. Our definition match with the relations defining the Quantum Group (equations (\ref{commutation-QG-1} ) (\ref{commutation-QG-2})( \ref{commutation-QG-2}) within the conventions adopted by Alvarez-Gaume et al. \cite{Alvarez-Gaume-QG}, while the definition of the deformation parameter in  \cite{Witten-Vertex-Models} agrees with the convention adopted in \cite{Pasquier}.
To make contact with the statistical Six-vertex model, one needs to take the Gauge group $G=SU(2)$ and compute the `Bob amplitude'. It was show using Skein theory that any four coupling (including over-crossing , under-crossing and pair creation) is equivalent to the Bob-amplitude :
\beq
A\ =\ u. \delta_{s_1,s_3} \delta_{s_2,s_4}+v\epsilon_{s_1,s_2} \epsilon _{s_3,s_4}. q^{-(s_1+s_3)} 
\eeq 
for some complex parameters $u, v$. Taking $u=(qx^{-1} -q^{-1}x)$ and $v= (x-x^{-1})$ the corresponding $R$-matrix $R_{s_i,s_j}$ is given by:
\bea
&& R(x,q)=(qx^{-1} -q^{-1}x) I+(x-x^{-1})U(q) \\ 
&& U(q)=\left[\begin{array}{cc}
          q & 1 \\
          1 & q^{-1}\\
\end{array}\right]\textit{ if $i< j$} \\
&& U(q)=0 \ ,  \textit{ if $i=j$} \eea
which is a possible form of the $R$-matrix for the Six-vertex model \cite{Pasquier}.
For a Gauge group $G=SU(2)$, the Chern-Simons vertex models exhibit symmetry $\widehat{U_q(sl(2))}$ in the same way that the classical Six-vertex model does. Therefore, the mathematical structure of the Quantum Groups encodes the topology of planar Wilson loops. A little of caution must be taken with the above defined Boltzmann Weights: first, note that since the Chern-Simons theory is well-defined for integer values of the coupling constant $k$, not all real values of the Boltzmann weights are well defined if we consider $q=\exp{(\pi i/k)}$. Second, the classical vertex model is defined in terms of classical degrees of freedom (the  currents). However,  the Chern-Simons action defines a non-dynamical theory ({\it i.e.}, there are no dynamical currents in CS theory). Both problems are solved by the following argument: let us impose boundary conditions to the EFT, {\it i.e.}, by compactifying the space domain onto a torus. Cutting down the torus along any cycle induces a loose of the Gauge symmetry, so that the Gauge fields become dynamical degrees of freedom, and as it is well-known that the CS theory becomes then equivalent to a chiral Weiss-Zumino-Witten (WZW) model defined on a circle \cite{Witten-2})
\beq
S_{CWZW}=-k \int_{CxR} d\theta dt tr[g^{-1}\de_\theta g g^{-1}\de_0 g  ] +\frac{k}{3}\int_{\Sigma x R} \epsilon^{\mu,\nu,\rho}
{\rm Tr}[[g^{-1}\de_\mu g g^{-1}\de_\nu g g^{-1}\de_\rho g]\ , \eeq 
where $g(z)$ is the chiral WZW field living in the group manifold of $G$. This is a conformal field theory (CFT) with central charge $c=(k+|SU(2)|)/(k+c_v)$, where $|SU(2)|$ is the number of generators of the $SU(2)$ Lie algebra, and $c_v$ is the dual coxeter number. The WZW model splits naturally into holomorphic and antiholomorphic pieces. The (holomorphic)  energy-momentum tensor is given by the Sugawara form \cite{Sugawara}:
\bea
T(z)=\frac{1}{2(k+c_v)}\sum_a :J_a(z) J_a(z):\ , \eea
where $k$ is the Chern-Simons coupling constant and $c_v$ is the dual coxeter number, which for $SU(2)$ is $c_v=2$. 
In other words, after quantization of the WZW model, there is a shift of the parameter $k \to k+c_v$ implied by the Sugawara construction.  The currents $J_a(z)$ and the tensor $T(z)$ may be expanded in modes in the usual way ( for a review see \cite{Di-Francesco}):
\bea
J_a(z)\sum_n z^{-n-1} J^a_n \\
T(z)=\sum_n z^{-n-2} L_nº . \eea
The WZW model has conformal and affine $\widehat{su(2)}_k$ symmetries  which can be written in terms of the 
Fourier modes of the Virasoro and current operators:
\bea 
&&[L_n,L_m]=(n-m)L_{n+m}+\frac{c}{12} \delta_{n+m,0}(n^3-n) \\
&& [J^a_n,J^b_m]=i\sqrt{2} \epsilon ^{abc} J^c_{n+m}+kn \delta^{ab} \delta_{n+m,0}\label {Kac-Moody-SU2}\\
&& [L_n,J^a_n]=-mJ^a_{n+m}\ . \eea
The identification between the CS theory and WZW model was used in \cite{Witten-Vertex-Models} to obtain an analytical continuation in $k$, which follows from the Knizhnik-Zamolodchikov  equation:
\bea
[\de_z-\frac{1}{k+g}\sum_{i\neq j}\frac{t_i^a t_j^a}{z_i-z_j}]\langle g(z_1,\bar{z}_1)....g(z_N,\bar{z}_N \rangle=0\ .
\eea
This can be used to compute the braiding matrices of this CFT. It has been shown that these braiding matrices correspond to the Boltzmann Weights of the associated 'interacting round a face '(IRF) model \cite{Alvarez-Gaume-QG}, which is equivalent to a  vertex model (this is the so-called face-vertex equivalence). 
The Boltzmann Weights can now be defined by the equations (\ref{Pesos-CS}), with the redefinition $q=\exp{(\pi i /(k+2))}$.
As it is known, the WZW model possesses a quantum group symmetry  $\uqa \otimes \uqa$ \cite{Alvarez-Gaume-QG}, with 
$q=\exp{(\pi i /(k+2))}$

\subsubsection{Abelian Chern-Simons Theory and Quantum Groups}\label{Abelian-CS-QG}

However, as it was pointed out by Witten \cite{Witten-Vertex-Models}, in the reduction from CS Gauge theory with Group $G$ to any vertex model, one losses the local and Global $G$ symmetry so that the vertex models retain only the maximal torus $T$ symmetry of the Group $G$. For the case at hand where $G=SU(2)$, the vertex model has a $U(1)$ symmetry which naturally corresponds to the Gauge symmetry of the charged fermions (charge density waves) propagating on the lattice. The extended  Kac-Moody symmetry $\widehat{su(2)}_1$ present in the WZW model is obtained by taking the equivalent CFT of (two) chiral bosons at self-dual point {\it i.e.}, where the currents:
\bea 
&&J^\pm =e^{\pm  i\sqrt{2}\phi}\\
&& J^z=i\de \phi \ , \eea
satisfy the  algebra $\widehat {su(2)_k}$  ( equation (\ref{Kac-Moody-SU2})) at level $k=1$. As we have seen in the previous Section, charge density waves are also naturally accounted by a {\it  Abelian} Chern-Simons theory on the circle. 

Imposing periodic boundary conditions on the square lattice amounts to  consider the Abelian CS theory on a (spatial) torus, such that the square lattice on the plane defines the homology cycles on the torus. In this domain, new degrees of freedom associated to the global Gauge transformations arise. To be more precise, let us consider the Lagrangian of the Abelian CS theory:
\beq
L=\frac{k}{4\pi} \int_{\bf T}  d^2x \epsilon _{ij}(\dot{A}_i A_j+A_oF_{ij})\ ,  
\eeq
where ${\bf T}$ denotes a torus with modular parameter $\tau$ and homology cycle basis $C_\a$ $C_\b$. As it is known (\cite{Nair-Bos-CS-CFT} \cite{Dunne-Review}), the Gauge field $A$ can be parametrized in this domain by using the Hodge decomposition, which  incorporates the  windings around the non-contractible loops on the torus. In holomorphic coordinates $z=x+ i y$ $\bar{z}=x-i y$ it is given by
\beq
A=\de _{\bar{z}} \chi +i\frac{\pi}{Img(\tau)} \bar{\omega}(z) a \ ,
\eeq
where the 1-form $\omega$ satisfies  $\int_{C_a} \omega =1$ and  $\int_{C_b} \omega =\tau$ and $a=a(t)$ is a complex (space independent ) function on time. The it can be shown that the Lagrangian, in the Gauge, $A_0=0$ becomes\cite{Dunne-Review}:
\bea
L_{cs}=i B_{eff}(\dot{a}a^* -\dot{a}^* a) +i k \int_{\sigma} (\de_{\bar{z}} \dot{\chi} \de_{z} \chi ^*-\de_z \dot{\chi}^*\de_{\bar{z}} \chi) \eea
The second term in the Lagrangian corresponds to Abelian CS  theory on the plane  with coupling constant $k$. The first one shows that the degree of freedom labeled by the function 'a' behaves as the coordinate of a quantum mechanical particle moving into an effective magnetic field $B_{eff}= \pi k /Img(\tau)$ restricted to the lowest Landau Level. 
The quantization of this theory is well known and can be done in the Sch{\"o}dinger picture \cite{Jacwiv-Dunne}. The wave functional is:
\beq
\Psi[A]=\psi(\chi)\psi(a) \ ,
\eeq
where $\psi(\chi)=e^{-\int \chi \frac{\de _+}{\de_-}\chi} e^{-\int|\chi|}$, and  $\chi=\sqrt {\frac{k}{4\pi}} (\chi_1+i\chi_2)$, $\de_{\pm}=(\de_1 \pm i\de_2)$. 
Small Gauge transformations do not affect the wave functional. However, due to the existence of non-contractible loops on the torus, the global Gauge transformations defined by \cite{Bos-Nair} :
\bea
&& a \rightarrow a+n_1+\tau n_2 \quad \chi \rightarrow \chi \\
&& A_{\bar z} \rightarrow -i\de_{\bar z} R R^{-1} \\
&& R(z,\bar{z})=exp(-\frac{n_1\pi}{Im\tau}\int^z_{z_0} (\bar{\omega}- \omega)  -\frac{n_2\pi}{Im\tau}\int^z_{z_0} \tau \bar{\omega}- (\bar{\tau}\omega)) 
\eea
affect both the zero and {\it a} modes. These large Gauge transformations are precisely the magnetic translations across a parallelogram unit cell:
\beq
T_R\psi(a)=e^{i \frac{B}{2} |a\wedge R|} \psi( a+ R)\ ,
\eeq
where:
\beq
T_R=e^{(\nabla +i e A) R} \qquad  T_a T_b= e^{i B/2 |a \wedge b| } T_{a+b}\ . \eeq
The exponential factor involves the flux through the parallelogram defined by $\vec{a}$ and $\vec{b}$. Now, 
following \cite{Sato} \cite{Kogan} we take the  combination of the magnetic translations :
\bea
&& E=\frac{1}{q-q^{-1}}[ T(a,a)-T(-a,a)] \\
&& F=\frac{1}{q-q^{-1}}[ T(-a,-a)-T(a,-a)] \\
&& K=T(a,0) \eea
where the translations are made from the elemental square plaquette of side $a$. These operators satisfy the relations:
\bea
[E,F]=\frac{K-K^{-1}}{q-q^{-1}} \\
kE=q^2EK \quad KF=q^{-2}FK \ , \eea
that define the Quantum Group $U_q(sl(2)$. Here $q=\exp(i B/2 a^2)=\exp(i\Phi/2)$ is the deformation parameter of the quantum group, and $\Phi$ is the flux per plaquette. Note that for $\Phi=2\pi$ we have $q=-1$ and then 
\bea
&& T_{(a,0)}\psi(x)=-\psi(x+a) \\
&& T_{(2a,0)} \psi(x)=\psi(x+2a)\ . \eea            
This is reminiscent to the staggered flux phase \cite{Afleck-Martson}, where the fluxes are antiferromagnetically ordered. Alternatively, the quantum group symmetry can  be also analyzed directly at the level of the Gauge transformations acting on the wave functionals and the resulting quantum group is also $U_q(sl2)$ where the deformation parameter is identified directly in terms of the coupling constant $k$ of the CS theory as $q=e^{i\pi/k}$ \cite{Grensing}.       
Summarizing, the Gauge invariance of the Chern-Simons theory on the torus implies the existence of a Quantum group symmetry  hidden in the theory. Now following \cite{Sierra-Book} we can define the generators:
\bea
&& E_0=e^uE  \quad  E_1=e^uE \\
&& F_0=e^{-u}E \quad  F_1=e^{-u}E \\  
&& K_0=K^{-1} \quad  K_1=K\ , \eea
where $x=e^u$ is an affinization parameter. These operators define a representation $(e^u ,1/2)$ of the affine Quantum Group $\widehat{U_q(sl2)}$. So that the double Abelian Chern-Simons Theory on the torus possesses a Quantum Group symmetry $\widehat{U_q(sl2)} \otimes \widehat {U_q(sl2)}$, which is identified with the quantum group symmetry of the Six-vertex model or, equivalently, with the symmetry Group of the two-dimensional lattice fermion model. To write down the corresponding EFT on the square lattice with periodic boundary conditions all we need to do is to take the modular parameter $\tau=i$ and to restrict the motion of the effective degrees of freedom (charge density waves) to the links on the lattice by replacing the  Chern-Simons term with the Lattice Chern-Simons term ( which posses lattice differential operators). Therefore, the EFT is
defined by the action:
\bea
S_{DCS}=\frac{k}{4\pi} \int d^3x\ a^R_\mu K_{\mu,\nu} a^R_{\nu} -\frac{k}{4\pi} \int d^3x\  a^L_\mu K_{\mu,\nu}a^L_{\nu}\ , \label{Double-CS}
\eea    
with $K_{\mu,\nu}=S_{mu}\epsilon_{\mu,\a,\nu}d_\a$, $S_\mu f(x)=f(x+a\epsilon_\mu)$, $d_\mu f(x)=(f(x+a\epsilon_\mu)-f(x))/a$, (where  $a$  is the lattice spacing), which  can also be written as a mixed Chern-Simons theory \cite{Carlo-Topics} \cite{Carlo-Supercond}.

\subsubsection{Identification of the Mott point ($q=-1$)}

We are now ready to identify the Mott point in the two-dimensional case. As we have seen in Section \ref{Integrability-2D-mott} the two dimensional Fermion model (\ref{Model-Ferm-2d}) which is equivalent to the $XXZ$ spin lattice, can be  split into one dimensional systems for each row or column of the square lattice , with Hamiltonian
\beq
H_{XXZ}^{1d}=\sum_{i=1}^{L}(S^x_iS^x _{i+1} + S^y_i S^y_{i+1}
-\frac{ q+q^{-1}}{2} S^z_i S^z_{i+1}) + H_{b}\ , \label{hxxz2} 
\eeq
where $H_{b}  = \a(S^z_1-S^z_L)$  and $\a=(q-q^{-1})/2$.

On the one hand the critical  point of this system, which has been analyzed in Section (\ref{section-1}), allows us to identify the Mott transition point in the square lattice with the  one-dimensional transition at $\Delta=1$ . Since we have defined $\Delta=-(q+q^{-1})/2$, it  implies q=-1. Besides the critical point in the one-dimensional system was identified with the WZW model or a Double Chern Simons theory on the circle.
On the other hand we have identified the critical point of the two dimensional (double periodic) lattice fermion system (\ref{Model-Ferm-2d})with the Double Chern Simon theory on the (space) torus with Quantum Group symmetry $\uqa \otimes \uqa $.  As we discussed in the Section (\ref{Vertex-model-CS}), cutting down the torus the Gauge field become dynamical currents represented by a WZW model, which naturally splits  its degrees of freedom in chiral components) with quantum group symmetry  $\uqa \otimes \uqa$ \cite{Alvarez-Gaume-QG} with $q=\exp{(i\pi/(k+2))}$.However, since the identification of the deformation parameter $q$ is done at the classical (Hamiltonian) level, there is no Sugawara shift in the WZW coupling constant $k$ due vacuum renormalization, an we should use $q=\exp{(i\pi/ k)}$.     
Therefore, since we can cut down the torus on any cycle, the affine Quantum group symmetry of the double Chern-Simons theory on the domain  $T \times R$  implies the quantum Group symmetry of each $XXZ$ spin system cut down in any circle.
We therefore conclude that $q=-1$ is the correct value of the deformation parameter characterizing the Mott point.

The above statement is the EFT formulation of the statement discussed in \cite{Bazhanov-1} and reviewed  in Section 3: the Quantum TE implies the standard Yang-Baxter equation, so that the integrability in the $(2+1)$-dimensional quantum system implies the integrability of the $(1+1)$-dimensional systems obtained from it by the projection-like character  of the solution (\ref{L-uniformizado}) .
Sumarizing, the EFT of the lattice Fermion system (\ref{Model-Ferm-2d})  at the Mott transition point is represented by a double Chern-Simons theory at coupling constant $k=1$,  in agreement with the analysis based on the magnetic algebra presented in the Section \ref{Vertex-model-CS}.

\subsection{The Order Parameter} 
  
We wold now like to discuss the emergence of the order parameter characterizing the transition. We can do this either from the point of view of the double-Chern-Simons theory or from the point of view of the equivalent CFT. 
First, we recall that, as we stated in the Section (\ref{Integrability-2D-mott}) , each row and column of the lattice system (\ref{Model-Ferm-2d}) can be considered as an one-dimensional system, which exhibit charge density wave order as we have stated in Section 1. At  $\Delta=1+\epsilon$ the one dimensional system in the row  is in gap-full state which correspond to the antiferromagnetic phase of the equivalent $XXZ$ spin chain. For $\Delta=1$, the one-dimensional fermion system is in a gapless phase and is described by a CFT with $\widehat{su(2)}_1\otimes \widehat{su(2)}_1$ symmetry, which is encoded in the WZW action  or in a chiral-antichiral bosonic system compactified on a circle at the self-dual radius. Let us now consider the transition in the square lattice: since we can cut along any cycle of the torus (defined in Subsection (\ref{Abelian-CS-QG}), we can obtain the above bosonic CFT at the self-dual radius in any row or column in the lattice.
Invariance under surgery of the states in the CS theory, implies by consistency that the fermion system on the lattice should be described by a CFT with $c=1$. This is interpreted as the theory of the free boson on the (space) torus, with partition function:
\beq
Z=\frac{1}{\eta(\tau)} \sum_{e,m}q^{1/2(e/R+m R/2)} \bar{q }^{1/2(e/R-m R/2)}\ , 
\eeq
where $\eta(\tau)$ is the Dedekind function, $\tau$ is the modular parameter of the torus and $e$ and $m$ are the electric and magnetic charges. Under $R$-duality ({\it i.e.}, the interchange $R \leftrightarrow 2/R$) the partition function is invariant if one exchanges the electric and magnetic charges ($e \leftrightarrow m$).\footnote{By including all possible boundary conditions (periodic and anti-periodic) one obtains the $S_1/Z_2$ orbifold partition function that is also related to the Six-vertex model at the critical line  \cite{Ginsparg}. We are not focusing in this case because we are working at half-filling states, which fixes the boundary conditions to be periodic}. 
This is also the partition function of an equivalent two-dimensional Coulomb gas described by the action:
\beq
S_{CG}=\frac{1}{2}\sum_{jk}[(\frac{e_j}{\sqrt{g}}+m_j\sqrt{g}) G(R_j-R_k)(\frac{e_k}{\sqrt{g}}+m_k\sqrt{g})]\ ,
 \eeq
where $G$ is the two-dimensional lattice Green function. Note that the above action describes the charge-charge , vortex-vortex and charge-vortex interactions. The Mott transition on the tours (square lattice with periodic boundary conditions) is given by the above partition function at the self-dual radius $R=\sqrt{2}$, such that the system exhibits electric-magnetic duality.

We would now like to discuss the behavior of the EFT away from the Mott critical point. 
We consider first the one-dimensional theory describing one of the orientations on the lattice. 
For $\Delta>1$ , the massive antiferromagnetic phase  is represented in the continuous limit by a Sine Gordon Theory:
\beq
S_{SG}=\int d^2x [\de_\mu \phi \de_\mu \phi + 2 \a_0 \cos \beta \phi]\ . 
\eeq
It is well-known that the partition function may be written as \cite{Amit-Sine-Gordon}: 
\beq
Z=\lim_{\epsilon \flecha 0} \sum_n \frac{\a ^{2n}}{(2n)!^2}\int \prod_{i=1}^{2n} d^2z_i \exp (\frac{\beta^2}{8\pi} \sum_{i\neq j} q_i q_j Ln|z_i-z_j|^2 +\epsilon^2) \ . 
\eeq
Here $z=x+iy$ denotes the position of the charges $q_i$  in complex coordinates,  that take values $\pm 1$. The  renormalized coupling constant $\a=\a_0 (\epsilon ^2)^{\frac{\beta^2}{8\pi}} $ plays the role of a  fugacity ( $\epsilon$ is a short distance regulator). This describes the antiferroelectric phase of the Coulomb gas model.

Moreover, the staggered  flux phase revealed by the quantum group analysis of the previous Section can be described by a CS theory defined on a torus with punctures, by defining pseudo-spins on the dual lattice  representing these fluxes.
If we define the dual system as the two-dimensional $XXZ$ model made from these pseudo-spins with coupling constant $\Delta '=\Delta ^{-1}$, then the self-dual point is defined by $\Delta=1$. Here the self-duality in defined by invariance under exchange among the spins with coupling constant $\Delta$ in the direct lattice, and spins in the dual lattice with coupling constant $\Delta'$. Just below the Mott transition ,{\it i.e.}, for $\Delta= 1-\epsilon$, the dual system  is in the frozen (Mott) state with the pseudo-spins in an Neel state, so that the fluxes are also  antiferromagnetically ordered. In this a case the EFT is given by a :
\bea
S_{CS}&=&\frac{k}{4\pi}\int d^3x\ a^\mu \ K_{\mu,\nu} a_\lambda + 
\sum^{'}_p \phi^0 \left[ \delta(x_d,y_d)\right . \nn\\
&-& \left . \delta(x_d+1,y)-\delta(x_d,y_d+1) + \delta(x_d+1,y_d+1) \right] \ ,
\label{CS-Vortex} 
\eea
where $a_\lambda$ is an Abelian CS field and $\sum^{'}_p$ means that 
the sum is taken over all fundamental domains. Each domain has period $2a$ and contains 
four vortices in antiferromagnetic array. The emergence of only one CS term reflects the breakdown of the  chiral  symmetry (as can be seen comparing the equation (\ref{CS-Vortex}) with the equation (\ref{Double-CS}) which contain two chiral fields ), and the classical low-lying states reproduce an staggered -flux phase current pattern . This fact coincides with the quantum group analysis at $q=-1$ presented in the Section (\ref{2D-EFT}). 
At the quantum level, Gauss law selects the physical states from the lattice CS Gauge theory on the torus with punctures. Therefore, the quantum order of the ground state of this theory is characterized as a staggered flux phase.


\vspace{1 cm}
The EFT that we have presented above allows one to study some properties of the system under doping. By analogy with the CS theory of the quantum Hall effect, we could expect a ground state stable against small doping. In that case, for the simplest inverse filling fractions $k=m$ (m odd integer), the ground state is described as a droplet of incompressible quantum liquid \cite{Laughlin-2} (however,  other phases with more exotic quantum orders, like Nematic phases are also possible in other regimes (\cite{Fradkin-Smectic-1}) (\cite{Fradkin-Smectic-2})  (\cite{Fradkin-Nematic}) and is stable under small perturbations away from the center of a given plateau in the conductivity. 
In the Mott system, we have already assumed that the dynamically generated vortices act as external statistical fields for the new electrons 
injected in the system by doping (this can be considered as an extension of the $R$-duality). At the self-dual point, statistical magnetic fields can be interchanged with statistical electric fields (on a torus). After imposing the lattice symmetries, the low-lying effective Hamiltonian for the injected electrons (in first quantization) is: 
\bea
H=\sum_i [-\hbar^2\frac{1}{2m}(\frac{\de^2}{\de x_i ^2} +\frac{\de^2}{\de y_i^2})+\lambda _i(x_i^2- y_i^2)] \ ,\label{effective-force} \eea
where $\lambda _i$ can take the values $\pm \lambda$.
Therefore, the electric potential changes sign in $x=\pm y$, producing domain walls between regions with different electron densities. Similar results can be obtained using the $W_4$ symmetry, which is related to the relevant perturbations of the Ashkin-Teller and Six-vertex models away from the critical point \cite{Bottesi-Zemba}\cite{Gaite}.      
            
One consequence of having discussed the EFT is that, {\it a-posteriori}, the behavior of the electrons can be more easily understood. 
It can be shown that the interaction term in the Hamiltonian (\ref{Model-Ferm-2d}) in the the continuum limit contains a
chemical potential term of the form $-\mu\ \rho$, with $\mu=\Delta$ , which ensures the half-filling condition. Therefore, changing  the chemical potential by doping in $\delta \mu$ modifies the Hamiltonian (in the spin representation (\ref{hxxz}) by:
\beq
 H(\Delta)\rightarrow H(\Delta)+\delta \mu \sum_{\langle i j\rangle} S^z_i S^z_j\ . 
\eeq
For $\Delta=1$,the dynamics of the electron system is given by the double CS theory (\ref{Double-CS}), whose Hamiltonian can be defined as the temporal component of the stress-energy tensor $H_{cs}=T_{00}$, where $T_{\mu\nu}=\delta S_{CS}/\delta g_{\mu\nu}$ and $g_{\mu\nu}$ is the metric tensor. However, the $CS$ action is topological and, therefore, independent of the metric implying $H_{cs}=0$
for each chiral component,  which leads to $H(\Delta=1)=0$. This means that doping the system away from the critical point, the dynamics is controlled by an effective Ising Hamiltonian.

\section{Conclusions}
 
In this paper we have studied the Mott transition in an interacting electron system with a hopping term and nearest neighbors density-density coupling defined on a square lattice (\ref{Model-Ferm-2d}) . We have first reviewed the one-dimensional case with periodic boundary conditions ( {\it i.e.} when the system lives on a circle) in Section 1, and we have written down a Conformal Field Theory description leading to the identification of the Mott transition point  
({\it i.e.}, when the coupling constant is $\Delta=1$) as the Lutter-Emery point in the bosonic formulation, and 
the degrees of freedom with charge density waves. We have also pointed out that the low-energy dynamics at the transition is described by a Wess-Zumino-Witten model, as it was already implicit in the literature.

We have also discussed the two-dimensional Mott transition, starting  with the study of the integrability of the two-dimensional fermion system (\ref{Model-Ferm-2d}). To do that, we have used  a two-dimensional Jordan-Wigner transformation and a new solution of the Zamolodchikov Tetrahedron Equation, which  allowed us to identify the Fermion system  (\ref{Model-Ferm-2d}) as a `Quantum vertex model', and shown that the fermion system is is characterized by an affine Quantum Group symmetry.   As a consequence of the projection-like property of the new solution, we concluded that the two-dimensional lattice system factorizes into two one-dimensional systems, one for any row and one for any column of the two-dimensional square lattice. This fact allowed for the identification of the two-dimensional Mott transition with the one-dimensional one, which occurs when the coupling parameter is  $\Delta=-(q+q^{-1})/{2}=1$. 
The identification of the symmetry and of the effective degrees of freedom (charge density waves) led to the 
construction of the Effective Field Theory at the critical point, whcih is a double (Abelian) Chern-Simons theory with Quantum Group symmetry $\uqa \otimes \uqa$ and deformation parameter given by $q=\exp{(-i\pi/k)}$. Furthermore, this effective theory may be considered as the broken phase of a non-Abelian Chern-Simons theory associated to the vertex models (which has been already pointed out by Witten).      

Finally, the behavior of the system near the Mott point has also been investigated using the ideas of EFT. We have found that the transition is of the Kosterlitz-Thouless class, characterized by
an array of Chern-Simons vortices in a anti-ferromagnetic order. This description corresponds to a $d$-density-wave order parameter for the matter currents. Upon doping with electrons, the magnetic-electric duality of the KT transition implies the appearance of domain walls between region of different densities.

 \def\RMP{{\it Rev. Mod. Phys.\ }}
 \def\PRL{{\it Phys. Rev. Lett.\ }}
 \def\PL{{\it Phys. Lett.}}
 \def\PR{{\it Phys. Rev.  \ }}
 \def\NP{{\it Nucl. Phys.}}
 \def\PRB{{\it Phys. Rev. B  \ }}
 \def\IJMP{{\it Int. J. Mod. Phys.}}
  
\end{document}